\documentclass[fleqn,usenatbib]{mnras}

\usepackage{newtxtext,newtxmath}
\usepackage[T1]{fontenc}

\DeclareRobustCommand{\VAN}[3]{#2}
\let\VANthebibliography\thebibliography
\def\thebibliography{\DeclareRobustCommand{\VAN}[3]{##3}\VANthebibliography}

\usepackage{graphicx}	
\usepackage{amsmath}	
\usepackage{amssymb}	
\usepackage{lscape}
\usepackage{rotating,booktabs}
\usepackage{arydshln}
\usepackage[section]{placeins}
\usepackage{caption}
\usepackage{float}
\usepackage{lipsum}
\usepackage{xcolor}
\newcommand{\zciv}{\hbox{$z_{\text{C\,{\sc iv}}}$}}
\newcommand{\lya}{Ly$\alpha$}
\newcommand{\HI}{\hbox{H\,{\sc i}}}

\newcommand{\HeII}{\hbox{He\,{\sc ii}}}

\newcommand{\Lya}{\mbox{Ly$\alpha$}}

\newcommand{\Hb}{\mbox{H$\beta$}}
\newcommand{\Ha}{\mbox{H$\alpha$}}

\newcommand{\kms}{\mbox{km s$^{{-}1}$}}
\newcommand{\cm}{cm$^{-2}$}
\newcommand{\ciii}{\hbox{C\,{\sc iii}}}
\newcommand{\civ}{\hbox{C\,{\sc iv}}}

\newcommand{\siiv}{\hbox{Si\,{\sc iv}}}

\newcommand{\mgii}{\hbox{Mg\,{\sc ii}}}
\newcommand{\oi}{\hbox{O\,{\sc i}}}
\newcommand{\oii}{\hbox{[O\,{\sc ii}]}}
\newcommand{\oiii}{\hbox{[O\,{\sc iii}]}}

\newcommand{\cii}{\hbox{C\,{\sc ii}}}

\newcommand{\Ociv}{\hbox{$\Omega_{\text{C\,{\sc iv}}}$}}

\newcommand{\ltsima}{$ \buildrel < \over \sim $}
\newcommand{\simlt}{\lower.5ex\hbox{\ltsima}}
\newcommand{\gtsima}{$ \buildrel > \over \sim $}
\newcommand{\simgt}{\lower.5ex\hbox{\gtsima}}

\title[Faint LAEs near \civ\ absorbers at $z>4.7$]{Faint LAEs near $z>4.7$
\civ\ absorbers revealed by MUSE\thanks{
Based on observations at VLT, ESO program 095.A-0714.
}}
\author[D\'{i}az et. al.]{
D\'{i}az, C. G.$^{1,2,3}$\thanks{E-mail:gdiaz@gemini.edu},
Ryan-Weber, E.V.$^{4,5}$,
Karman, W.$^{6}$,
Caputi, K.I.$^{6}$, 
\newauthor  Salvadori, S.$^{7,8}$, Crighton, N.H.$^{4}$, Ouchi, M.$^{9,10}$, 
and Vanzella, E.$^{11}$\\
$^{1}$Gemini Observatory, Southern Operations Center, La Serena, Chile \\
$^{2}$Instituto de Ciencias Astron\'omicas, de la Tierra y del Espacio (ICATE), San Juan, Argentina\\
$^{3}$Consejo de Investigaciones Cient\'ificas y T\'ecnicas (CONICET), San Juan, Argentina\\
$^{4}$Centre for Astrophysics and Supercomputing, Swinburne University of Technology, Hawthorn, VIC, Australia\\
$^{5}$ARC Centre of Excellence for All Sky Astrophysics in 3 Dimensions (ASTRO 3D)\\
$^{6}$Kapteyn Astronomical Institute, University of Groningen, Groningen, The Netherlands \\
$^{7}$Dipartimento di Fisica e Astronomia, Universita di Firenze, via G. Sansone 1, Sesto Fiorentino, Italy\\
$^{8}$Istituto Nazionale di Astrofisica (INAF) - Osservatorio Astrofisico di Arcetri, Largo E. Fermi 5, Firenze, Italy\\
$^{9}$Institute for Cosmic Ray Research, The University of Tokyo, Kashiwa, Japan\\
$^{10}$Kavli Institute for the Physics and Mathematics of the Universe (WPI), The University of Tokyo, Kashiwa, Japan\\
$^{11}$INAF - Osservatorio di Astrofisica e Scienza dello Spazio di Bologna, via
Piero Gobetti 93/3, I-40129 Bologna, Italy}

\date{Accepted XXX. Received YYY; in original form ZZZ}

\pubyear{2020}

\begin{document}
\label{firstpage}
\pagerange{\pageref{firstpage}--\pageref{lastpage}}
\maketitle

\begin{abstract}
We present the results
from the search for Lyman Alpha emitters (LAEs)
in the proximity of 11 \civ\ absorption systems at $z>4.7$
in the spectrum of the QSO J1030+0524,
using data from MUSE.
We have found multiple LAE candidates 
close to four \civ\ systems at $z_{\text{\civ}}=4.94$--5.74
with $\log_{10}(N_{\text{\civ}}[$\cm$])>13.5$.
At $z=5$--6, \civ\ systems with W$_0(\text{\civ})> 0.2$ \AA\
seem more likely to have galaxies with \Lya\ emission 
within $\rho< 200$ proper kpc (4/5 cases), than the \civ\ systems 
with W$_0(\text{\civ})<0.2$ \AA\ (0/6 cases).  
The impact parameter of LAE-\civ\ systems with
equivalent widths $W_0($\civ$) > 0.5$ \AA\ 
is in the range $11\, \simlt\,\rho\,\simlt\,200$ proper kpc (pkpc).
Furthermore, all candidates are in the luminosity range
0.18--1.15 L$^{\star}_{\text{\Lya}}(z=5.7)$,
indicating that the environment of \civ\ systems 
within 200 pkpc is populated by the faint end of 
the \Lya\ luminosity function.
We report a 0.28\,L$^{\star}_{\text{\Lya}}$
galaxy at a separation of $\rho=11$ pkpc from a 
strong \civ\ absorption ($\log_{10}(N_{\text{\civ}}[$\cm$])=14.52$)
at $z_{\text{\civ}}=5.72419$.
The prevalence of sub-L$^{\star}_{\text{\Lya}}$ galaxies
in the proximity of $z>4.9$ \civ\ systems
suggest that the absorbing material 
is rather young, likely ejected 
in the recent past of the galaxies at these redshifts.
The connection between faint LAEs
and high-ionization absorption systems reported in this work,
is potentially a consequence of the role of low mass galaxies in the early
evolution of the circum-galactic  
and intergalactic media.
\end{abstract}

\section{Introduction}\label{s:intro}

About 10 percent of the baryon content of the Universe 
is accounted for by condensed visible matter in galaxies, groups and clusters.
The remaining large majority of baryons exist in a gaseous state outside of galaxies
\citep[e.g.][]{persic1992, shull2012}, 
covering a large range of temperatures and densities,
from the gas in the multi-phase circum-galactic medium (CGM)
within a few virial radii,
to the diffuse intergalactic medium (IGM)
filling the vast space between galaxies
\citep{ferrara2005, prochaska2011, tejos2014, peeples2014, werk2014, lehner2014, wakker2015, wotta2016}.

In the current cosmological framework of galaxy formation,
cold streams of baryons from the IGM flow into galaxies
providing the raw material for star-formation 
\citep{keres2005, dekel2009, borthakur2015, nielsen2013}. 
However, only a fraction of the inflowing gas can form
stars because star formation
itself, via supernova explosions and stellar winds,
introduces large amounts of energy and momentum to the 
interstellar medium (ISM) {\bf by} removing 
large quantities of gas
from the inner regions of the galaxies
resulting in galactic winds or outflows
 \citep{oppenheimer2008, hopkins2012, muratov2015}.
This source of mechanical feedback 
contributes regulating star-formation and
is required by hydrodynamical simulations
and semi-analytic models to reproduce 
several observational results such as: the 
star formation rate (SFR) and galaxy stellar mass functions 
\citep{oppenheimer2010, hopkins2014, somerville2015}, 
the fraction of gas and galaxy metallicities \citep{dave2011b},
the luminosity functions in the rest-frame UV and optical \citep{fontanot2017b},
and even the observed properties of present-day ancient dwarf galaxies 
dwelling in the Local Group \citep{salvadori2015, revaz2018}.
Moreover, outflows can
introduce metals in the CGM and the IGM
\citep{oppenheimer2009, cen2011,pallottini2014},
which are commonly  observed as metal absorption line systems 
in the spectra of background light sources like high redshift quasars (QSOs).

Metal absorption systems provide a wealth of information
about the absorbing gas, including velocity, covering fraction, 
metallicity, density, temperature and ionization state.
In recent years, observations have revealed the presence of gas 
accretion \citep{rubin2012, bowen2016, ho2017}
and outflows of gas from high redshift star-forming galaxies
\citep{pettini2001,steidel2010, bradshaw2013, karman2014, rubin2014}. 
Triply ionized carbon (\civ) has now been detected in intervening systems out to redshifts $z\sim6.5$ \citep{ryan-weber2009, bosman2017, codoreanu2018, meyer2019, cooper2019}. 
The presence of \civ\ has been reported in the CGM of Lyman Break Galaxies (LBGs) at $z=2-3$, with $\log_{10}(N_{\text{\civ}}[$\cm$])>14$ 
within $\sim 90$ proper kpc (pkpc) from the closest galaxy 
\citep{steidel2010, turner2014}
and tend to be commonly detected in regions of galaxy
overdensity \citep[][]{adelberger2005b}. 
Low redshift studies of \civ\ absorbers find a close association 
with galaxies out to $\sim 250$\,pkpc \citep{chen2001, stocke2013, ford2014, bordoloi2014}. 
In a series of papers \citet{burchett2013,burchett2015, burchett2016} 
conducted a search for faint galaxies associated with 
low redshift ($z<0.015$) \civ\ systems and explore 
the relationship between individual galaxies as well as 
their environments. They found that \civ\ was preferentially 
associated with $M_{\star}>10^{9.5} M_{\odot}$ galaxies 
in low density environments.
Although these observations demonstrate a connection between
galaxies, the CGM and the IGM,
we are just starting to understand the relation between
the physical conditions in the ISM
and the conditions of the gas outside of galaxies.

Several studies based on cosmological hydrodynamical 
simulations have found that the evolution of low ionization 
metal absorptions, like \oi\ and \cii\
are relatively insensitive to the choice of galactic feedback, 
whereas high ionization species like \civ\ and \siiv\ are highly
sensitive to the feedback prescription \citep{tescari2011, keating2016, rahmati2016, garcia2017b}.
This is particularly important at high redshift where
metal systems are possible tracers of galaxies hiding 
below current detection limits \citep{becker2015b}.

At the end of the epoch of hydrogen reionization ($z\sim6$)
the ionizing ultra-violet background (UVB) is
predicted to have large spatial fluctuation in intensity 
and spectral slope \citep[e.g.][]{mesinger2009,finlator2015,finlator2016}.
The non-uniform spatial distribution of the ionizing sources
results in a mean free path of ionizing photons 
that varies with density.
The inclusion of these UVB fluctuations in simulated 
data improves the match to the observed statistics
of high ionization metal ions like \civ\ and \siiv\ at $z>5$
\citep{oppenheimer2009, finlator2016}.
In this scenario, the UVB is enhanced in environments
dominated by ionizing sources. 
Therefore, studying the connection between 
galaxies and high ionization metal absorptions like \civ\
at $z>5$ can give us critical information about 
the production of metals and ionizing photons in
such environments.

In three previous publications we have reported on the search
for galaxies with the 
hydrogen Lyman-$\alpha$ line in emission
(``Lyman Alpha Emitters'' or LAEs) 
to characterize the environment of \civ\ systems at $z\sim5.7$.
The field of the QSO J1030+0524 was chosen because
it contains 11 \civ\ absorption systems between $4.7<z<6.1$
revealed by high-resolution spectroscopy \citep{dodorico2013}. 
The strongest system known in this redshift range at $z_{abs}\sim 5.724$ \citep{ryan-weber2006}
inhabits an over-density of LAEs on scales of 10 comoving Mpc \citep{diaz2014}
and has a galaxy counterpart at $\sim213$ pkpc \citep{diaz2015}.
This large distance to the \civ\ system is in tension with:
(i) observations of the $z\sim2$--3 CGM \citep{steidel2010, turner2014},
(ii) predictions from cosmological simulations \citep{oppenheimer2009, keating2016},
and (iii) the typical wind speeds measured on star-forming galaxies 
\citep[]{shapley2003,bradshaw2013,hashimoto2013b,karman2014}.

A simple explanation is that there are additional galaxies  
below the detection limit of \citet{diaz2014}.
In favor of this idea, recent simulations modeling high redshift 
intergalactic absorption systems from \citet{garcia2017a}
suggest that dwarf galaxies ($-20.5<$\,M$_{\text{UV}}<-18.8$ mag)
are responsible for the metal absorptions observed at $z\sim5.7$. 
These luminosities are fainter than the galaxy sample of \citet{diaz2014}.
In particular, \citet{garcia2017a} 
reports that dwarf galaxies (M$_{\star}\sim 2 \times10^9$M$_{\odot}$) 
with a $\sim100$\kms\ wind speed
could be the type of source  
associated to the \civ\ systems
observed at $z_{\text{\civ}}= 5.72419$.

In this work, we have deepened our search for galaxies 
in the field of the QSO J1030+0524
using the Multi-Unit Spectroscopic Explorer (MUSE) on the Very Large Telescope (VLT). 
MUSE offers both the area and sensitivity to search for sub-L$^{\star}_{\text{\Lya}}$ 
galaxies that could be responsible for the \civ\ absorptions.
The main goals of this study are: {\it a)} to determine if there is a fainter 
galaxy closer to the strong \civ\ at  $z_{\text{\civ}}= 5.72419$ than the LAE
from \citet{diaz2015}, and {\it b)} to search for galaxies near 
the other 10 \civ\ systems 
in the QSO's line of sight.
The first objective aims to address 
the hypothesis that faint galaxies 
closer to the line of sight are responsible for metals absorbers at $z>5$.
The second objective will contribute to
the expansion of the sample of $z>5$ galaxy-\civ\ system pairs that are required to reconstruct the history of baryons across cosmic time.

Observations are described in Section \ref{s:obs}.
A detailed explanation of the detection method 
can be found in Section \ref{s:detection},
and the resulting LAEs
are presented in Section \ref{s:results}. 
The discussion in Section \ref{s:discussion} 
reviews the the connection between 
star-forming galaxies and high-ionization
absorption systems at high redshift.
The conclusions can be found in Section \ref{s:conclusion}.
In this work we use {\it Plank 2014} cosmology 
\citep[$H_{0}=67.79 \pm 0.78$\,\kms Mpc$^{-1}$,
$\Omega_{\text{M}}= 0.308 \pm 0.010$ and 
$\Omega_{\Lambda}=0.692\pm 0.010$,][]{plank2014}.

\section{Observations}\label{s:obs}
QSO J1030+0524 was targeted with the MUSE 
\citep[][]{bacon2012} for 2 hours on April 10 2015 
and 6 hours between January 7 2016 and 
January 10 2016\footnote{ESO programme 095.A-0714, PI Karman}. 
The conditions during observation were good, 
with a seeing of $\sim1\arcsec$ reported during 
the first two hours and a seeing better than 
0.7$\arcsec$ for the remaining 6 hours. 
The pixel scale is 0.2 arcsec/pixel
and the spectral resolution is 1.25 \AA/pixel.

Each pointing employed the same observing strategy, 
where we set up two observation blocks of 1440 seconds
which followed a dither pattern with offsets of a fraction of 
arcsecond and rotations of 90 degrees
to better remove cosmic rays and to obtain a better noise map.  
The total observation time of QSO J1030+0524 with MUSE, 
correcting for overheads, amounted to 6.4 hours.

We followed the data reduction as described in 
\citet{karman2015} for both pointings, and 
refer to that paper for details. Here we provide 
only a brief description of the data reduction.
We used the standard pipeline of MUSE Data 
Reduction Software version 1.0 on all of the raw data.
This pipeline includes the standard reduction 
steps like bias subtraction, flat-fielding, wavelength
calibration, illumination correction, and cosmic ray removal. 
The pipeline combines the processed raw data frames
 into a data-cube that includes the variance of every pixel 
 at every wavelength. We subtracted the remainder 
 of the sky at every wavelength in the obtained 
 data-cube by measuring the median offset in 11 
 blank areas at every wavelength, and subtracting
this from the entire field. 
The full width at half maximum (FWHM) was measured for 
the QSO and a second point-like object in the field, at various wavelengths.
We confirmed the excellent 
observing conditions with 
a FWHM of $\sim0.7\arcsec$, though
slightly higher at the shorter wavelengths.

The field of QSO J1030+0524 was previously
observed with the Advanced Camera for Surveys (ACS)
on the \textit{Hubble Space Telescope (HST)}
in the i'-band (F775W) and the z'-band (F850LP) \citep{stiavelli2005}.
A description of the HST data reduction can be found in \citet{diaz2011},
although in this work we use these archival images for visual inspection only.
We note, however, that none of the sources detected with MUSE
was previously identified by studies of high redshift galaxies
based on these HST images like \citet{stiavelli2005} and \citet{kim2009}.
As we will show in Section \ref{s:candidates},
most LAEs from the present work do not have 
a robust counterpart in the HST images.
Indeed, these images are not deep enough 
to detect with sufficient significant the faint galaxies reported here. 

\section{Detection of \Lya\ emission line}\label{s:detection}

The search for emission line galaxies was carried out
in three different ways focusing on the redshifts around the \civ\ absorbers. We have not conducted a full blind search of the entire MUSE cube.
The first sample of candidates was obtained
by visual inspection (VI) 
as described in section \ref{s:vi}.
Then, the automatic detection (AD) described in 
section \ref{s:ad} returned a second independent 
sample of LAE candidates, 
which largely overlaps with the sample from VI.
Finally, a third detection process
was developed to search for sources 
in narrow-band images that contain emission 
lines only. We refer to this technique as
``Differential Image Detection'' (DID), 
which is described in section \ref{s:did}.
This last procedure is more reliable
so it was used for confirmation 
of candidates from the other two techniques.

\subsection{Visual inspection (VI)}\label{s:vi}

The field of view (FoV) was divided in nine 
smaller sections and the data cube was 
inspected, frame by frame, in wavelength windows 
of 200 \AA\ centered at the wavelength 
corresponding to \Lya\ (1215.668 \AA)
at the redshift of each of the 11 \civ\ absorption systems 
in Table \ref{t:civ}, reported in \citet[][Table A3 therein]{dodorico2013}.  
As a result, each of the nine small fields
was surveyed in 11 windows in wavelength. 
The wavelength windows 
of 200 \AA\ corresponds to $\Delta v \sim 7400$\,\kms,
which largely exceeds 
the redshift range in which a galaxy 
could be physically associated to 
the corresponding \civ\ system.

We searched for bright objects covering 
several spatial pixels that remain
detectable in at least three consecutive frames
(in the wavelength direction). 
Then we extracted a spectrum 
using a 2\arcsec\ aperture (diameter). 
This was compared with 10 sky spectra 
obtained from regions of the FoV
that have no objects to rule out sky residuals.
As a result, five LAEs were identified 
with this technique.
The next section describes the analysis of the data-cube 
with an independent automatic detection tool. 

\begin{table}
	\caption{\civ\ systems in the spectrum of QSO J1030+0524
		from \citet{dodorico2013}. Columns are: (1) reference number,
		(2) redshift, (3) equivalent width of the \civ\ doublet, and 
		(4) column density.}
	\label{t:civ}
	\begin{tabular}{llcc}
		\hline
		\multicolumn{1}{c}{CIV} &
		\multicolumn{1}{c}{$z(\text{\civ})$} &
		\multicolumn{1}{c}{W$_0(1548,1550)$} &
		\multicolumn{1}{c}{$\log_{10}(N_{\text{\civ}})$}\\
		\multicolumn{1}{c}{ID} &
		\multicolumn{1}{c}{(redshift)} &
		\multicolumn{1}{c}{(\AA)} &
		\multicolumn{1}{c}{(\cm)} \\
		\hline  
		1&    4.76671 & 0.139 &  13.13 $\pm$ 0.03\\
		2&    4.7966  & 0.105 &  13.30 $\pm$ 0.04\\
		3&    4.79931 & 0.2   &  13.37 $\pm$ 0.02\\
		4&    4.80107 & 0.526 &  13.46 $\pm$ 0.01\\
		5&    4.89066 & 0.119 & 13.21 $\pm$ 0.02\\
		6$^a$&    4.9482 & 0.49  &  13.22 $\pm$ 0.04 \\
		&     &    &  13.77 $\pm$ 0.01 \\  
		7$^a$&    5.5172 & 0.61  & 13.4 $\pm$ 0.2  \\  
		&      &   &  13.92 $\pm$ 0.05 \\  
		8&    5.72419 & 1.24  &  14.52 $\pm$ 0.08 \\  
		9$^a$&    5.7428 & 0.79  &  13.8 $\pm$ 0.1  \\  
		&       &   & 13.89 $\pm$ 0.09 \\   
		10&    5.9757  & 0.07  & 13.1 $\pm$ 0.3  \\    
		11&    5.9784  & 0.15  & 13.4 $\pm$ 0.2  \\ 
		\hline\end{tabular}
	\\
	$^a$Two components \civ\ absorption.
\end{table}

\subsection{Automatic Detection (AD)}\label{s:ad}

The data-cube was analyzed with the 
{\it MUSE Python Data Analysis Framework}
\citep[MPDAF\footnote{http://mpdaf.readthedocs.io} 2.1,][]{bacon2016}. 
For each \civ\ absorption system, 
a narrow wavelength section of 100 \AA\ around
the observed wavelength of \Lya\ was extracted from the cube. 
This small cube was scanned with MUSELET 
(MUSE Line Emission Tracker), which runs 
an automatic search for line-emission objects. 
The process creates narrow-band images and runs SExtractor 
\citep{bertin1996} to detect isolated emission lines
and emission line sources with continuum, 
using the default aperture size of 1.6'' (diameter).
Default configuration parameters
and a gaussian PSF convolution mask with FWHM$ = 2.0$ pixels
were used for source detection.

The output list of emission line sources with 
no continuum detection was sorted in velocity
respect to the \civ\ to identify the closest candidates
in the line-of-sight direction. 
Then, low redshift contaminants were removed 
based on the detection 
of other emission lines, as described below in 
section \ref{s:contaminants}.
This technique recovers three of the five LAEs 
from Visual Inspection and two new LAEs.

\subsection{Differential Image Detection (DID)}\label{s:did}

This section describes a procedure to highlight 
emission line objects in the field of view, by removing all other sources
using narrow-band (NB) images.
For every \civ\ system in Table \ref{t:civ}, the process was the following. 
A small cube of 10 \AA\ ($\sim50$--75\,\kms\ rest-frame)
with central wavelength
corresponding to \Lya\ at the redshift of the \civ\ absorption,
was collapsed in wavelength direction 
using a variance-weighted sum with weights $w_i=1/\sigma_i^2$
for the $i-$th pixel,
to create a NB image corresponding to rest-frame \Lya.
The \Lya\ forest (Lyf) and the UV continuum (UV) were sampled with
two small (10 \AA) cubes shifted 40 \AA\ to the blue and
40 \AA\ to the red of \Lya, respectively.
These two cubes were collapsed 
to produce NB images of the \Lya\ forest
and the UV continuum. No filter transmission was applied.
Then, one NB image was subtracted from another to produce 
a differential image (DI).
The following DIs were calculated: 
(NB(\Lya)-NB($\text{UV}$)) and (NB(\Lya)-NB(\Lya$_{\text{forest}}$)),
in which most sources are removed while the sources
with flux excess in the NB(\Lya) 
(e.g. with an emission line) are revealed.

\begin{figure*}
	\includegraphics[scale=0.6]{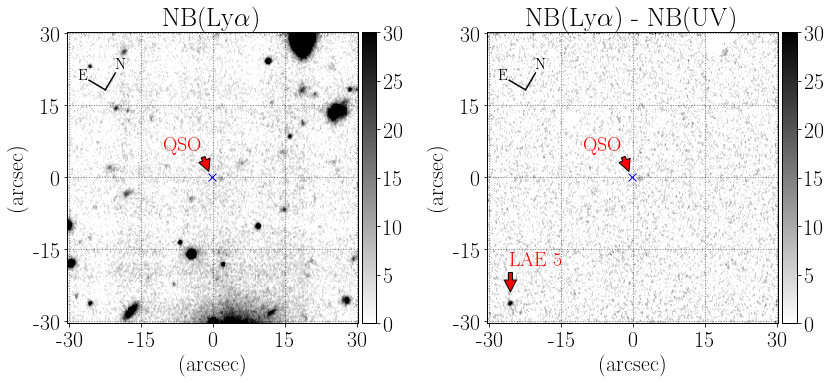}
	\caption{ LAE \#5 in MUSE's field of view.
		The NB image on the left has a central wavelength corresponding
		to \Lya\ at $z=5.721$ 
		and the image on the right is a differential image 
		NB(\Lya)-NB(UV).}
	\label{f:9-fov}
\end{figure*}

Figure \ref{f:9-fov} presents the FoV 
of MUSE with QSO J1030+0524 at the center.
The image on the left is 
centered at $\lambda\sim8176$\,\AA,
which is the wavelength of \Lya\ at $z\sim5.725$.
It has tens of sources, some of them are obvious (bright)
low-redshift and galactic sources, but most of them are faint sources.
The image on the right panel is (NB(\Lya)-NB(${\text{UV}}$)),
where the UV continuum 
was subtracted from the \Lya\ line.
In this image, most sources in the field 
are completely removed, 
while LAE $\#5$ (see Table \ref{t:candidates})
is clearly visible. 

DIs were thoroughly examined 
to detect all possible sources
both visually and with SExtractor.
The list of SExtractor detections 
was obtained using default
configuration parameters except for 
{\it DETECT\_MINAREA = 3 pixels},
{\it DETECT\_THRESH = 1.5} 
and {\it SEEING\_FWHM=0.7}.
Each detection was analyzed to remove 
lower redshift interlopers. 
After this process, only one or two LAE candidates
were left per \civ\ system.
Overall, this method returns six sources (LAEs $\#1$, 2, 3, 5 and 6), 
and only one of them (LAE $\#6$)
was not detected with other methods.

In addition, DID was used to test 
other candidates from AD and VI 
that lie at $\sim 800$ \kms\ from 
a \civ\ absorption (LAEs 4, 7 and 8).
The analysis confirmed these three detections
and demonstrates the agreement between the techniques.
Therefore, the final sample of eight LAEs
in Table \ref{t:candidates} is considered  consistently detected by the DID technique.

\subsection{Removal of low redshift contaminants}\label{s:contaminants}

For every emission line detection, the spectrum
was thoroughly inspected for additional emission 
lines that would reveal a low redshift 
contaminant. In this exercise, it was assumed that the 
line was \Ha, \Hb, \oiii(5007\AA), \oiii(4958\AA), 
{\bf or} \oii(3727\AA). For each case, the wavelength of other 
optical emission lines was calculated. Then, we examined
the cube (3D) and the spectrum (1D) to asses the presence 
of additional emission lines. In this process, 
several emitters of \oiii, \oii, and \Ha\ 
have been identified at 
$z= 0.33$, 0.42, 0.567, 0.825, 1.01, 1.03 and 1.2. 
The sources that show no other emission line are 
the LAEs in the sample of Table \ref{t:candidates}. 

\section{Results}\label{s:results}

\subsection{LAE candidates}\label{s:lae1-images}

\begin{figure}
	\includegraphics[scale=0.5]{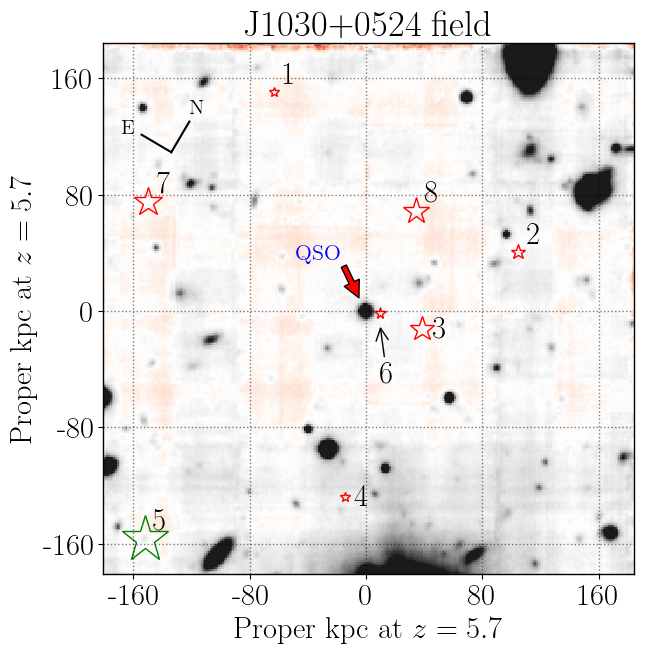}
	\caption{Position of the LAEs of Table \ref{t:candidates} 
		in MUSE's FoV. The background QSO
		is at the center of the field, 
		the red and green star symbols indicate the positions of the
        LAEs and the size of the symbol is proportional to the \Lya\ line flux.}
	\label{f:pos-all}
\end{figure}

We report the detection of 8 LAE candidates in 
the datacube.
The positions are indicated in Figure \ref{f:pos-all}
and presented in Table \ref{t:candidates},
along with the redshift,
angular distance to the QSO l.o.s.,
 impact parameter and radial velocity to the \civ\
 system, the total line flux, \Lya\ luminosity,
rest-frame equivalent width (EW$_0$)
of the emission line (assuming it is \Lya) and
the skewness of the line profile.

\begin{figure*}
	\includegraphics[scale=0.5]{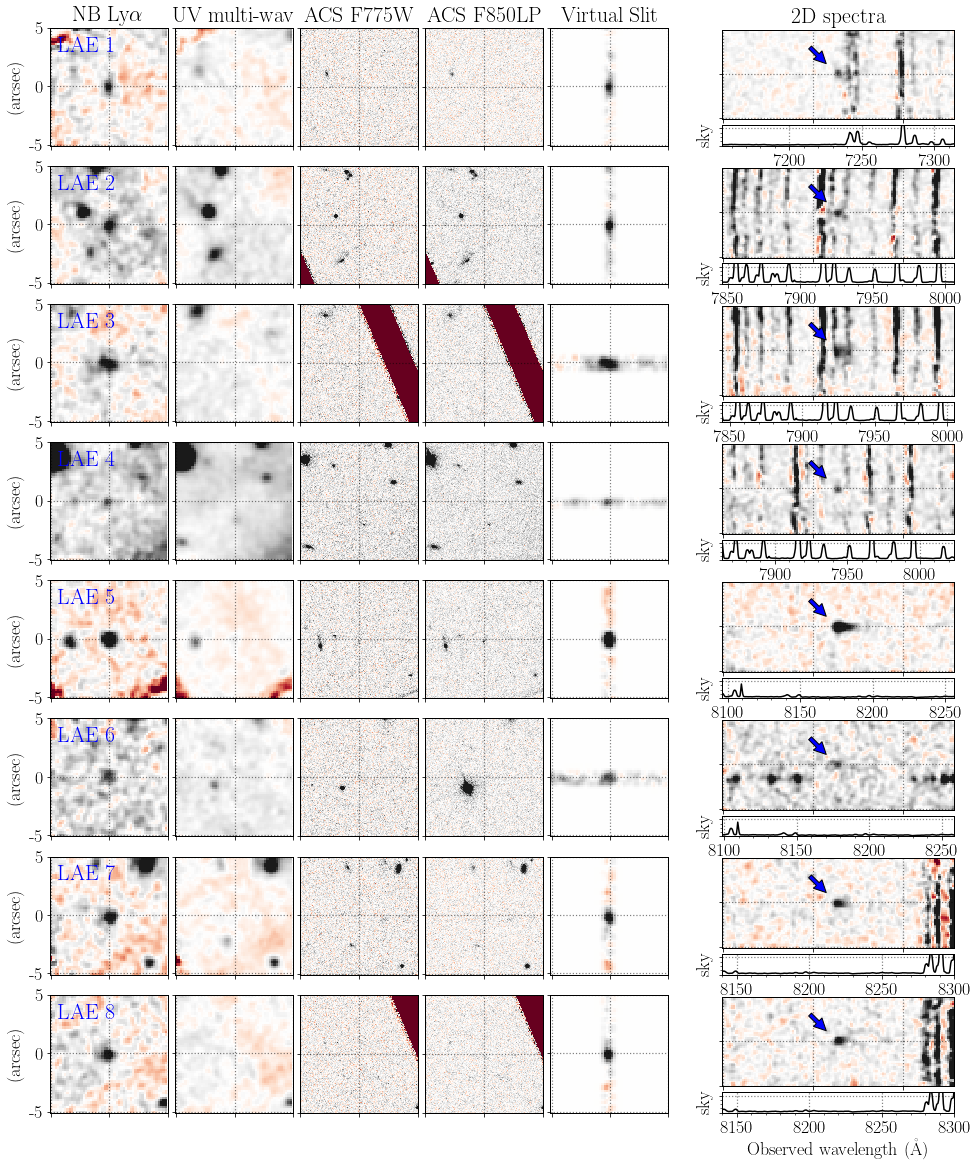}
	\caption{LAE candidates. 
		Images from left to right are: NB(\Lya),
		UV continuum extracted from
		multiple wavelength windows,
		i'-band (ACS/HST), z'-band (ACS/HST), 
		the slit used for extraction of the 2D spectrum,
		and the 2D spectrum with an arrow indicating
		the position of the emission line and a
		sub-plot showing the emission spectrum of the sky
		for reference.}
	\label{f:laes_1}
\end{figure*}

Figure \ref{f:laes_1}
presents a closer look at each galaxy
with thumbnails of 10\arcsec\ by 10\arcsec ,
centered at the position 
of the corresponding detection,
and the 2D spectrum extracted from 
the data-cube.
The first two columns from left to right,
NB(\Lya) and UV continuum,
are images extracted from the data-cube.
They sample the \Lya\ emission 
and the UV continuum between
1220\AA\ to 1300\AA. The latter was 
obtained from multiple 
wavelength windows defined to avoid
strong sky emission residuals
which are clearly visible in the 2D spectrum
of most sources.
The third and fourth columns are ACS/HST
images in the i' and z' bands, respectively.
Only LAE \#5 can be identified in 
the z'-band image ACS/HST.
We ran SExtractor several times
using different detection configurations
and weight types, and we found
that none of the other sources 
have detectable z'-band counterparts.

The emission lines were confirmed
in the 2D and 1D spectra.
Figure \ref{f:laes_1} shows the 2D spectra
extracted using virtual slits
of 0.7 to 1\arcsec\ width, which are displayed in the same figure.
The emission lines are indicated with 
blue arrows.

\begin{table*}
	\begin{minipage}{170mm} 
		\caption{LAEs from the search with MUSE.
			Columns are: (1) \civ\ system identification,
			(2) LAE reference number,
			(3) right ascension hh:mm:ss.ss (J2000), 
			(4) declination $\pm$dd:mm:ss.s (J2000),
			(5) redshift assuming the blue edge of the line profile is 1215.668 \AA,
			(6) angular distance to the QSO in arcseconds,
			(7) transversal distance \civ\ absorption in proper kpc,
        	(8) line of sight velocity to the \civ\ absorption in \kms,
			(9) total line flux in erg s$^{-1}$ \cm, 
			(10) luminosity of the emission line in erg s$^{-1}$ assuming it is \Lya,
			(11) lower limits on the rest-frame equivalent width in \AA,  and
			(12) weighted skewness.} 
		\label{t:candidates}
		\begin{tabular}{cccllrrccccc}
			\hline
			\multicolumn{1}{c}{System} &
			\multicolumn{1}{c}{LAE\#} &
			\multicolumn{1}{c}{RA} &
			\multicolumn{1}{c}{Dec} &
			\multicolumn{1}{c}{z$(\text{\Lya})$} &
			\multicolumn{1}{c}{$\delta\theta$} &
			\multicolumn{1}{c}{$\rho$} &
			\multicolumn{1}{c}{$\Delta V$} &
			\multicolumn{1}{c}{F$_{\text{\Lya}}/10^{-18}$} &
			\multicolumn{1}{c}{L$_{\text{\Lya}}/10^{41}$  } &
			\multicolumn{1}{c}{EW$_{0}$} &
			\multicolumn{1}{c}{S$_w$} \\
			\hline
			\hline  
	\civ\,6 & 1 & 10:30:28.56 & +05:25:11.4 &4.947 & 27.242 &176&71 &3.1$\,\pm\,$0.4&   8.3$\,\pm\,$1.1  & $>$14&-1.4$\,\pm\,$0.8       \\
			\hline  
 	\civ\,7 & 2 & 10:30:26.32 & +05:25:09.7 &5.518 & 18.672 &114&2  &4.4$\,\pm\,$0.5&   15.4$\,\pm\,$1.7  & $>$ 4    & 0.6$\,\pm\,$0.6    \\
		        &	3 & 10:30:26.64 & +05:24:56.4 &5.518 &  6.807&42&2  &7.4$\,\pm\,$0.5&   25.5$\,\pm\,$1.6  & $>$ 8 &  2.0$\,\pm\,$0.6    \\
	             &	4 & 10:30:26.52 & +05:24:35.4 &5.530 & 21.419  &131&-566 &2.8$\,\pm\,$0.4&    9.5$\,\pm\,$1.4  & $>$ 4   & -2.7$\,\pm\,$0.9 \\ 
			\hline  
  	\civ\,8  &  5 & 10:30:27.68 & +05:24:19.8 &5.721 & 36.411 &218& 127 &14.4$\,\pm\,$0.3&   52.6$\,\pm\,$1.1  & $>$ 44  &4.3$\,\pm\,$0.7    \\
		       &  6 & 10:30:26.99 & +05:24:56.1 & 5.720 &  1.627 &10&172 &3.4$\,\pm\,$0.3&    12.7$\,\pm\,$1.1  & $>$8  & -1.9$\,\pm\,$1.6    \\
			\hline  
	 	\civ\,9  & 7 & 10:30:28.96 & +05:24:53.1 &5.758 & 27.887  &167&-639&8.5$\,\pm\,$0.3&   32.1$\,\pm\,$1.1 & $>$27    & 4.0$\,\pm\,$1.1  \\
	 	&	8 & 10:30:27.15 & +05:25:07.9 &5.758 & 12.854  &77&-639&8.0$\,\pm\,$0.3&   30.4$\,\pm\,$1.1  & $>$15 & 8.0$\,\pm\,$1.5    \\
			\hline
		\end{tabular}
	\end{minipage}
\end{table*}


The 1D spectra were extracted using 
circular apertures of diameter equal to
$2\times$ the FWHM
of the sources measured by SExtractor 
in the detection image.
The 1D spectrum and
the aperture of extraction
of each object 
are presented in Figures \ref{f:ap1} to \ref{f:ap8}.
The comparison sky background
is represented with a cyan-filled area.
It was extracted from an annular aperture
centred on the source,
with a 2\arcsec\ inner radius and a 5\arcsec\ outer radius.
All foreground objects where masked out
before computing the mean flux density 
in the sky aperture.

The emission redshift reported in Table \ref{t:candidates}
was determined from 
the blue edge of the emission line profile
assuming that most of the line flux is
heavily absorbed by neutral hydrogen in the emitting galaxy
\citep[e.g.][]{verhamme2008}.
The velocity difference between the LAE and the \civ\
system is calculated as:
\begin{equation}
\Delta V(\text{\civ - Ly}\alpha)=(1- \frac{ \lambda_{obs}}{1215.668 \times (1+z_{\text{\civ}})})c 
\end{equation}
where $\lambda_{obs} $ is the observed wavelength of the emission line,
$z_{\text{\civ}}$ is the redshift of the corresponding \civ\ system,
and $c$ is the speed of light in the vacuum.
The impact parameter (or transversal distance) in proper kpc 
is obtained as:
$\rho=\delta \theta/g(z_{\text{\Lya}})$
where $\delta \theta$ is the angular separation
between the corresponding LAE and the QSO,
and $g(z_{\text{\Lya}})$ is the scale factor 
in \arcsec/pkpc at the redshift of the \Lya\ emission.

The weighted skewness (S$_w$) of the line profile,
which quantifies the asymmetry of the emission line,
was measured as defined in \citep{kashikawa2006b}.
Five LAE candidates have positive skewness
which is associated  with the absorption by neutral hydrogen
on the blue side of the line profile.
The remaining three candidates, those with
negative skewness, are the faintest sources in the sample.
In particular, the large error on S$_w$ for LAE \#6
results from a faint emission line whose symmetry is 
affected by noise (1D spectrum in Figure \ref{f:ap6}).
The other two objects with negative skewness are
LAE \#1 and LAE \#4. The effect of including these
two candidates in the analysis will be considered in
Section \ref{s:discussion}.

The total line flux is calculated from the 
integral over the line profile, which is
indicated in the central panels of 
Figures \ref{f:ap1} to \ref{f:ap8}
with a grey-shaded area between green lines.
The integral was approximated by 
a weighted sum of the form:

\begin{equation}
\label{eq:FxA}
\text{F}_{{\rm Ly\alpha}} =   \frac{\sum  f_i  \, w_i \, \Delta\lambda} {\sum w_i}\times n_{\text pix}.
\end{equation}
where $f_i$ is the flux density of the $i$-th pixel,
$w_i$ are the weights calculated from the variance spectrum
as  $w_i = 1/\sigma_i^2$, and $\Delta\lambda=1.25$ \AA$/$pix 
is the spectral resolution. 
The errors reported in Table \ref{t:candidates}
are estimated from:
\begin{equation}
\label{eq:FxAer}
\delta\text{F}_{{\rm Ly\alpha}} =   \frac{\sum w_i^2} {(\sum w_i)^{2}}\times 10^{-18}.
\end{equation}

In all candidates, the continuum flux density
redder than \Lya\ is an upper limit
because is below the detection limit of the data
given by the error spectrum,
or at least comparable to the flux level 
measured in the sky aperture.
The continuum
was estimated as the average of 
several windows in wavelength  
to avoid high skylines residuals the best we could,
and for each window the mean flux density
was obtained from a weighted sum:
\begin{equation}
\label{eq:fdens}
\text{f}_{\text {UV}} <  \langle \frac{\sum_{i}  f_{i,k}  \, w_{i,k} } {\sum_{i} w_{i,k} } \rangle,
\end{equation}
where $f_{i,k}$ and $w_{i,k}$ are 
the flux density and the weight of the $i$-th pixel and the 
$k$-th window. 

Finally, the lower limits on the rest-frame \Lya\ equivalent width (EW$_0$)
are obtained as:
\begin{equation}
\label{eq:EW0}
\text{EW}_{0} > \frac{\text{F}_{{\rm Ly\alpha}} }{\text{f}_{\text UV} } \times \frac{1}{(1+z)}.
\end{equation}

\subsection{Individual LAEs}\label{s:candidates}

\begin{figure}
	\includegraphics[width=80mm]{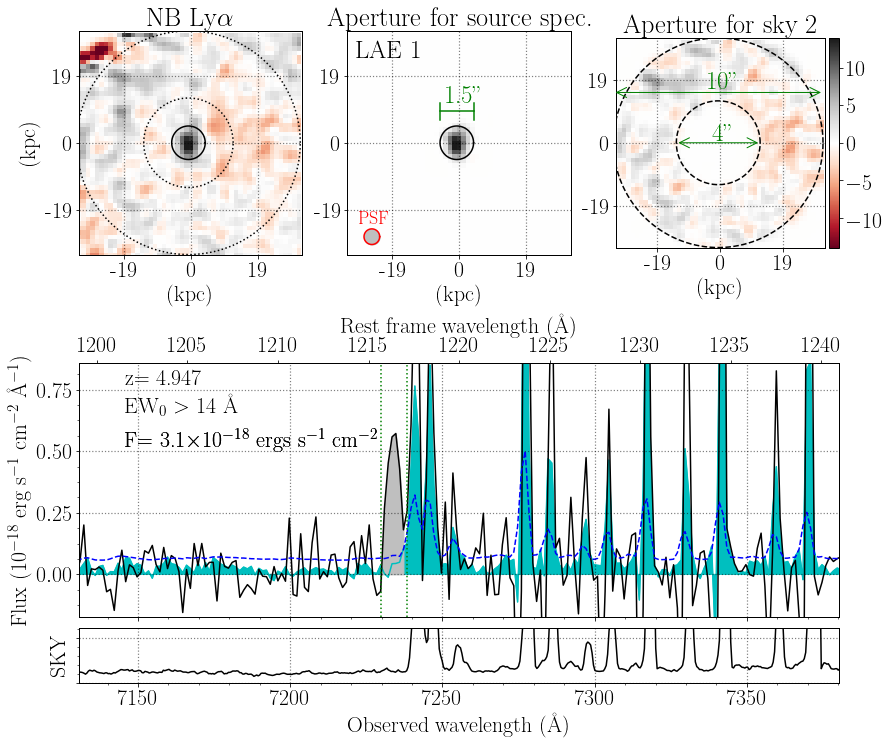}
	\caption{{\bf LAE $\#1$.} The top figures show 
		the	NB(\Lya) image (left), the aperture used for extraction of the 1D
	spectrum (center) and the aperture used for the extraction of 
	the sky spectrum (right) for comparison.
    The middle panel shows the 1D spectrum of the LAE
     with a black line and the 1D spectrum of the sky with
     a cyan-filled area. The \Lya\ emission line is indicated
     by the grey-filled area between dotted green vertical lines.
     The error spectrum is the blue dashed line.}
	\label{f:ap1}
\end{figure}

\begin{figure}
	\includegraphics[width=80mm]{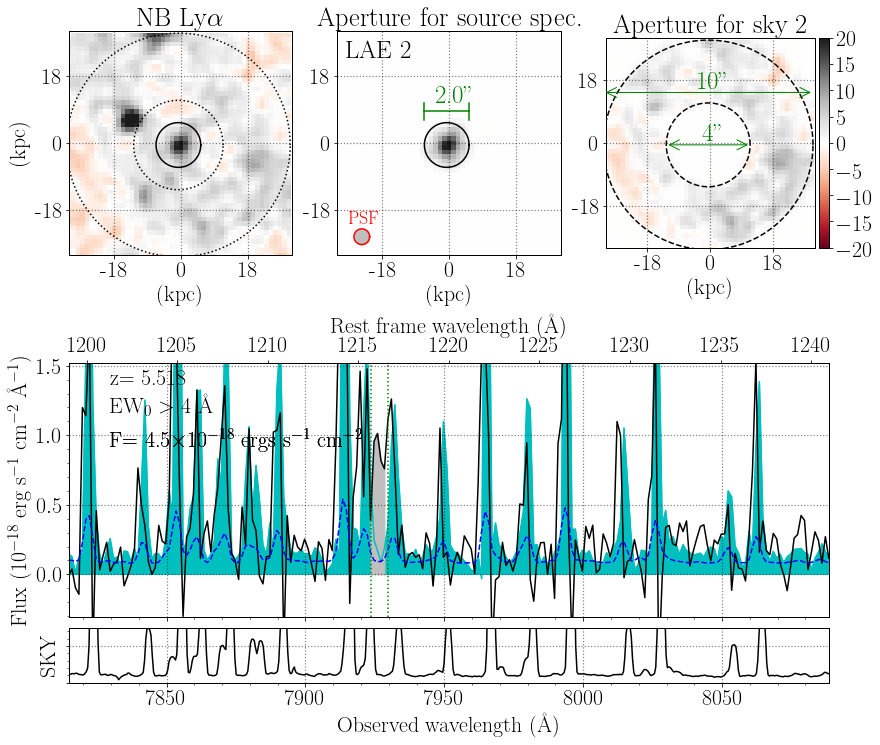}
	\caption{{\bf LAE $\#2$.} Description as per Figure \ref{f:ap1}.}
	\label{f:ap2}
\end{figure}

\begin{figure}
	\includegraphics[width=80mm]{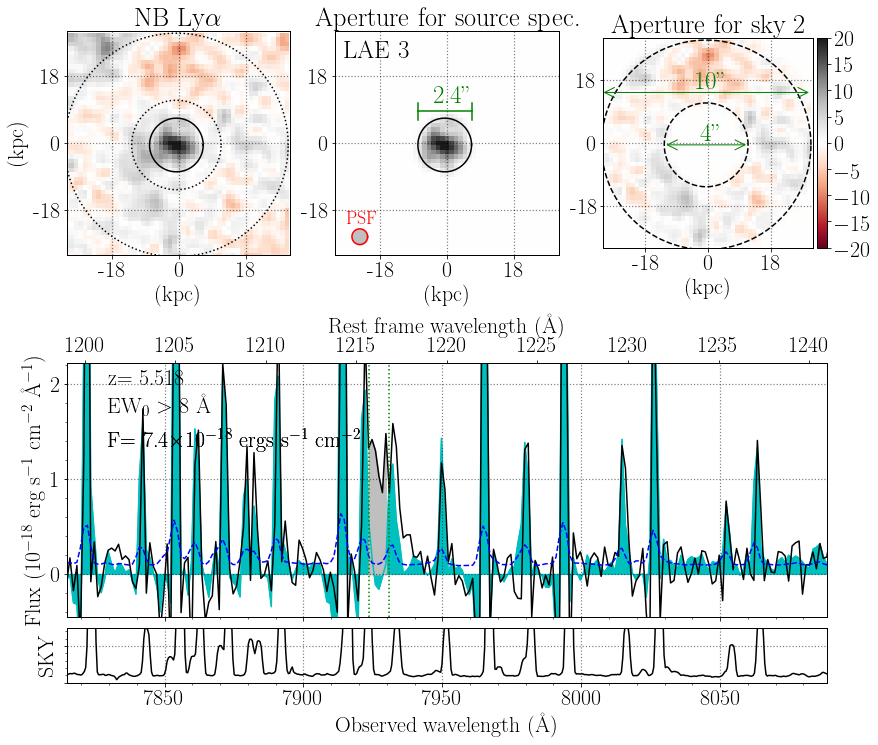}
	\caption{{\bf LAE $\#3$.} Description as per Figure \ref{f:ap1}.}
	\label{f:ap3}
\end{figure}

\begin{figure}
	\includegraphics[width=80mm]{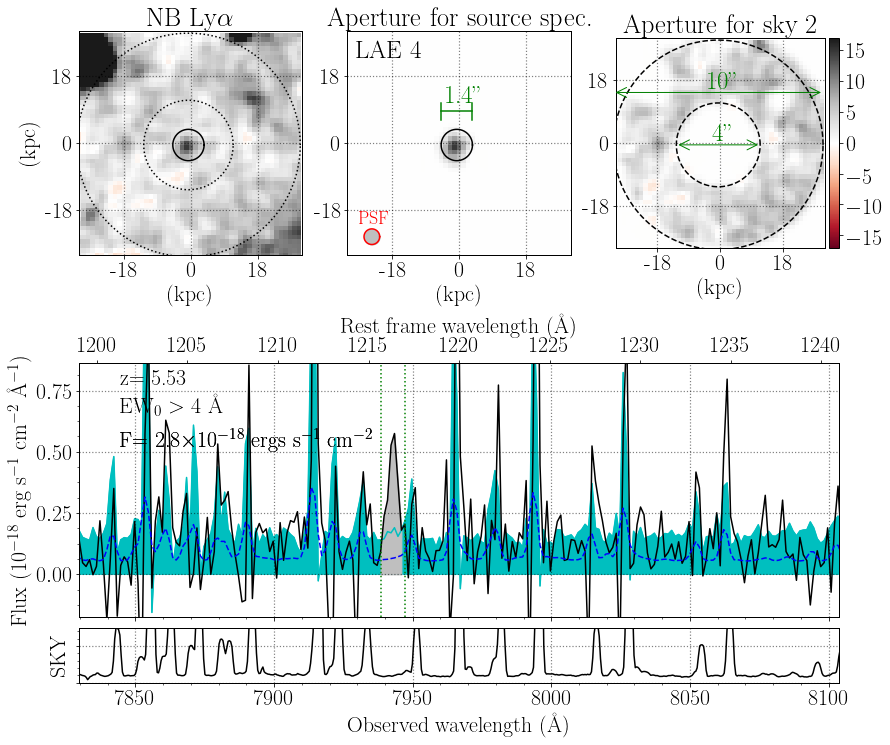}
	\caption{{\bf LAE $\#4$.} Description as per Figure \ref{f:ap1}.}
	\label{f:ap4}
\end{figure}

\begin{figure}
	\includegraphics[width=80mm]{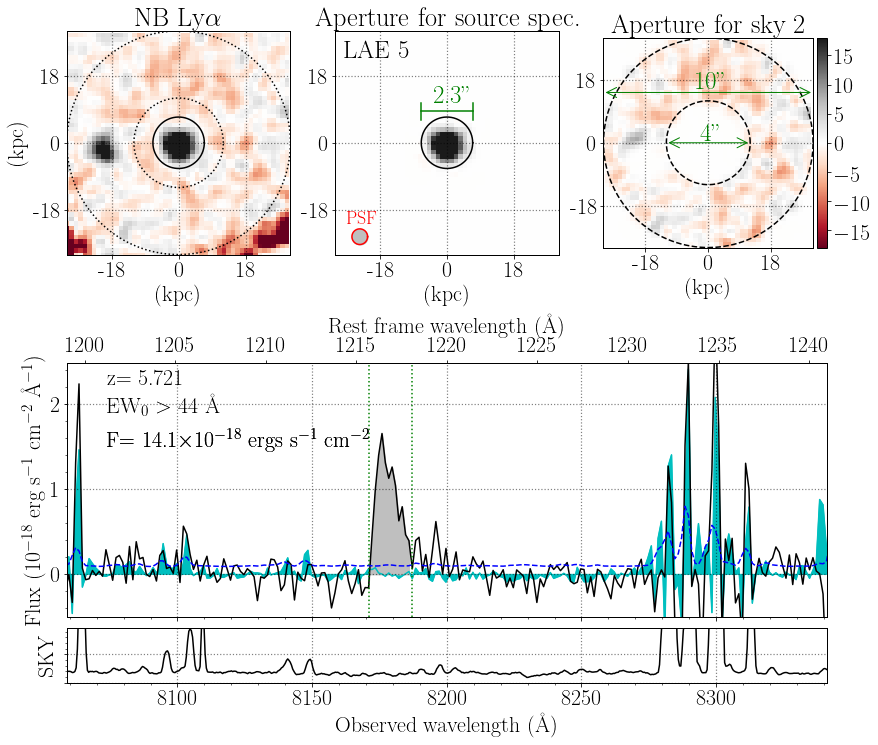}
	\caption{\small {\bf Spectrum of LAE 5.} Description as per Figure \ref{f:ap1}.}
	\label{f:ap5}
\end{figure}

\begin{figure}
	\includegraphics[width=80mm]{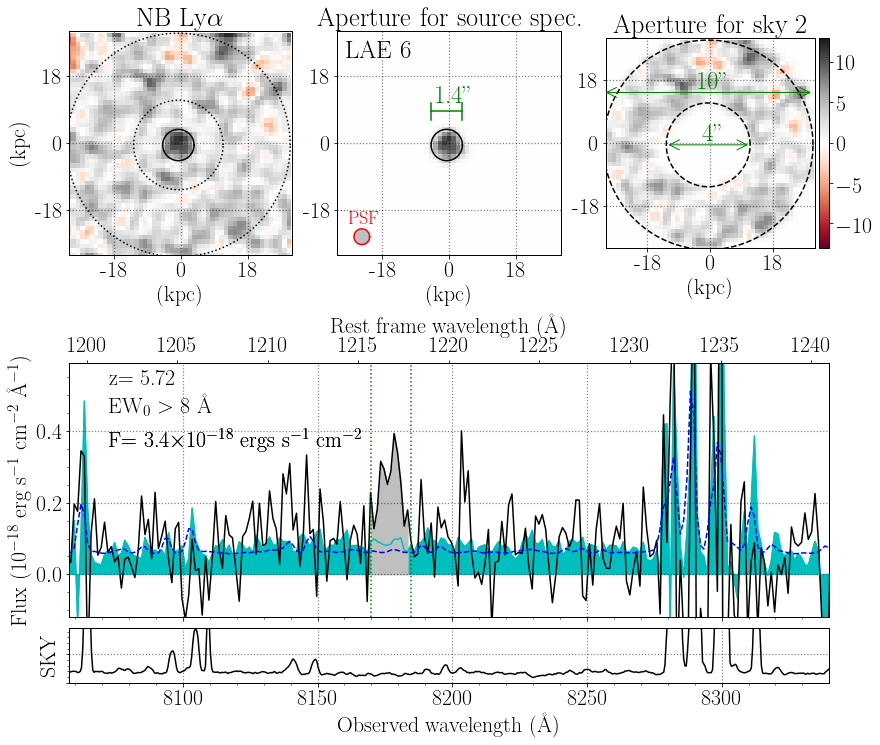}
	\caption{{\bf LAE $\#6$.} Description as per Figure \ref{f:ap1}.}
	\label{f:ap6}
\end{figure}

\begin{figure}
	\includegraphics[width=80mm]{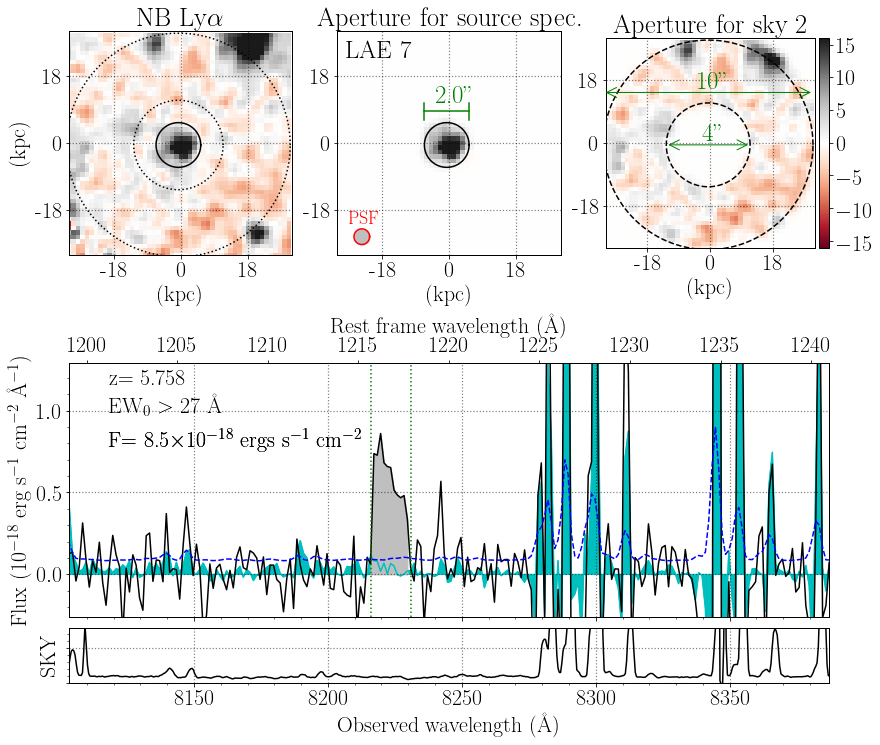}
	\caption{{\bf LAE $\#7$.} Description as per Figure \ref{f:ap1}.}
	\label{f:ap7}
\end{figure}

\begin{figure}
	\includegraphics[width=80mm]{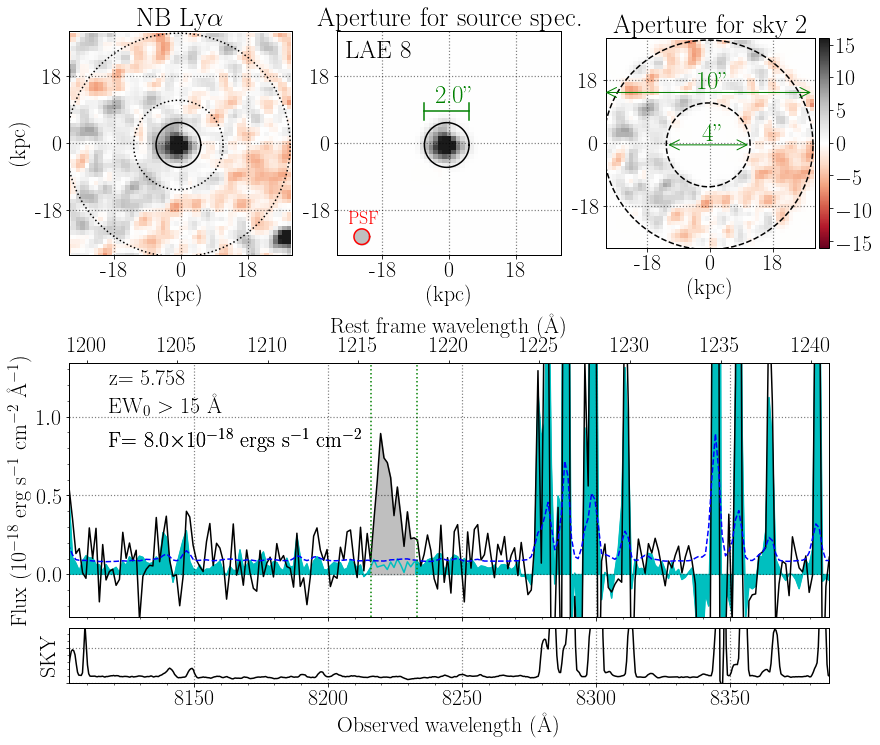}
	\caption{{\bf LAE $\#8$.} Description as per Figure \ref{f:ap1}.}
	\label{f:ap8}
\end{figure}

Figures \ref{f:ap1} to \ref{f:ap8}
present three thumbnails of 10 arcsec wide
in the top row,
centered on the corresponding candidates:
the NB(\Lya) image (left panel),
the aperture for extraction of the 1D spectrum
of the source (middle panel)
and the aperture of extraction of the sky
for comparison (right panel). 
The center panel presents the 1D spectrum
obtained from the source aperture (top middle)
with solid black line. The error spectrum
is represented by the blue dashed line
and the 1D spectrum from the sky aperture (top right)
is presented as filled cyan spectrum.
The vertical dotted lines indicate the limits
of the line profile considered to compute the total
flux reported in Table \ref{t:candidates}.
Finally, in the bottom panel we present
the sky spectrum for reference, in an arbitrary scale.

\subsubsection{LAE 1}\label{s:lae1}

This candidate was detected 
at 27.242\arcsec\ from the QSO
and is the only LAE for the \civ\ system 
at $z_{\text{\civ}}=4.9482$.
The emission line is detected 
at $z_{\text{\civ}}=4.947$,
which puts this object at 176 pkpc and
$\Delta V=70.8$\,\kms\, from the \civ\ system.
The 1D spectrum of Figure \ref{f:ap1}
shows a faint emission line 
that is absent in the comparison 
sky spectrum. 
The skewness of the line profile 
is negative, but it should be noted
that a fraction of the object emission line 
could be coincidental with the strong sky 
emission line at $\lambda=7240$\,\AA\,
and thus not included in the calculation of the 
line flux.

\subsubsection{LAE 2}\label{s:lae2}

This object lies at 18.672\arcsec\ 
from the QSO sight-line
with an emission redshift of $z_{\text{BLUE}}=5.518$,
which puts it at 114 pkpc from the \civ\ system at 
$z_{\text{\civ}}=5.5172$.
The emission is observed 
between bright sky emission lines
(Figure \ref{f:ap2}),
therefore the redshift estimate is limited
and the asymmetry of the line profile cannot 
be measured with high confidence.

\subsubsection{LAE 3}\label{s:lae3}

LAE \#3 is also at $z=5.518$,
making it a neighbor of LAE \#2.
It lies at 6.807\arcsec\ (42 pkpc) 
from the QSO line of sight,
thus it is closer to the 
\civ\ system at $z_{\text{\civ}}=5.5172$
than LAE \#2.
In the close view in Figure \ref{f:laes_1},
the object is clearly seen in NB(\Lya)
but there is no evidence in the rest-frame UV.
The 2D spectrum shows an emission
line brighter than LAE \#2
but similarly affected by the sky emission.

\subsubsection{LAE 4}\label{s:lae4}

LAE \#4 is a faint source at
21.419\arcsec\ from the QSO sight-line.
If it is \Lya\, the redshift is $z_{\text{BLUE}}=5.530$
and the impact parameter is 131 pkcp.
Thus, it is near the \civ\ system at $z_{\text{\civ}}=5.5172$.
Figure \ref{f:laes_1} shows a faint detection in NB(\Lya) 
and nothing in the other images. 
The emission line is confirmed in the 2D 
and 1D spectra (Figure \ref{f:ap4}).
The velocity difference to the \civ\ system 
is lager than 500 \kms, and the asymmetry of the line is 
negative which could be indicative
of a lower redshift interloper.

\subsubsection{LAE 5}\label{s:lae5}

This is the brightest LAE in the sample.
It was previously reported by \citet{diaz2014}
and confirmed in \citet{diaz2015}.
In this work, we measure an angular distance to the QSO line of sight 
of 36.411\arcsec\ and a redshift of $z_{\text{\Lya}}=5.721$.
As a result, it lies at 
218 pkcp from to the strong \civ\ system at $z_{\text{\civ}}=5.72419$.
LAE \#5 is clearly seen in the z'-band image in Figure \ref{f:laes_1}
and, like all other source in the sample, LAE \#5 is detected in NB(\Lya)
but not detected in the i'-band, which is typical of high-$z$ galaxies.
The 2D and 1D spectra are the highest signal-to-noise 
of the sample and the asymmetry of the line 
is obvious in both representations (Figure \ref{f:ap5}).
Line flux measurements, redshift, S$_w$ and 
EW$_0$ estimates are in agreement 
with previously reported values.

\subsubsection{LAE 6}\label{s:lae6}

This LAE candidate is at 1.627\arcsec\ from the QSO
line of sight.
The emission line is consistent with \Lya\ at $z_{\text{BLUE}}=5.720$
that would make it the closest LAE to the \civ\ 
system at $z_{\text{\civ}}=5.72419$, and a satellite of LAE \#5.
The object is faint in NB(\Lya) and there is no trace 
of it in other images.
The virtual slit in Figure \ref{f:laes_1} captures 
the LAE and part of the QSO light. As a result,
the 2D spectrum shows the emission line of LAE \#6
and the \Lya\ forest in the QSO spectrum.
The 1D spectrum in Figure \ref{f:ap6} shows a 
faint emission line in a region free of strong sky residuals.
The signal is spread across several 
adjacent pixels in the sky (e.g. NB(\Lya)) 
and in wavelength direction (2D and 1D spectra).
The low signal-to-noise of the emission line,
which seems quite symmetric,
results in large errors in S$_w$. 
Thus, asymmetry should not be used 
to evaluate the nature of the emission line.

\subsubsection{LAE 7}\label{s:lae7}

This galaxy is a solid detection,
the asymmetry of the line profile is consistent with 
high-$z$ \Lya\ at $z_{\text{BLUE}}=5.758$,
which is at $\Delta V\sim -638.6$\,\kms\ 
to the \civ\ system at $z_{\text{\civ}}=5.7428$.
LAE \#7 is at 27.887\arcsec\ (167 pkpc)
to the East of the QSO sight-line
and it is only detected in NB(\Lya) 
like most of the sample.
The emission line is not contaminated by sky residuals,
and the 2D and 1D spectra clearly show an asymmetric profile
confirmed by S$_w$ (Table \ref{t:candidates}).

\subsubsection{LAE 8}\label{s:lae8}

This LAE is at the same redshift of LAE \#7 ($z_{\text{BLUE}}=5.758$)
and the line profile supports the LAE identification.
It is at 12.854\arcsec\ (77 pkpc) to the North of the QSO sight-line,
thus it is the closest LAE to the \civ\ system at $z_{\text{\civ}}=5.7428$.
The emission in the 2D spectrum 
is typical of high-$z$ LAEs and the 1D spectrum confirms 
the asymmetry of the line
with a skewness of S$_w=8.0\pm1.5$.

\subsection{ Comparison with previous studies of the same field}\label{s:previous}

The field of the QSO J1030+0524 has been 
studied before with different instruments.
Here we use MUSE data to review the LAE candidates
identified by previous works.
Figure \ref{f:cai1} shows MUSE's FoV,
the position of the LAE candidates in Table \ref{t:candidates},
the position of the two LAE candidates in \citet{cai2017}
with blue circles, the position of the candidates
from \citet{diaz2011} with blue diamonds,
and the LAE previously confirmed by \citet{diaz2015}.

\begin{figure}
	\includegraphics[scale=0.5]{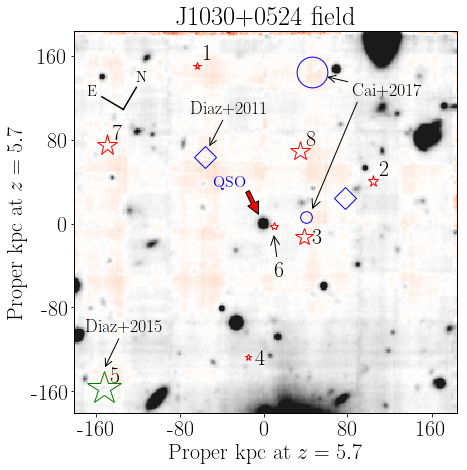}
	\caption{
		Same as Figure \ref{f:pos-all}
		including the two detections from \citet{cai2017} (circles)
		and two from \citet{diaz2011} (diamonds).The circle size is scaled to the flux 
		expected for the \Lya\ luminosity reported in \citet{cai2017}.}
	\label{f:pos-others}
\end{figure}

\subsubsection{\citet{diaz2011} }\label{s:diaz11}

\begin{figure*}
	\includegraphics[scale=0.45]{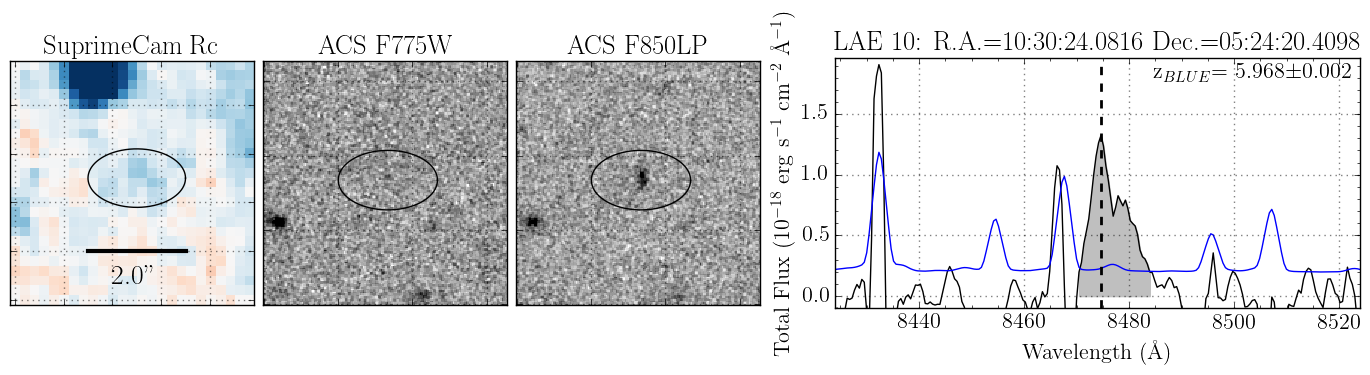}
	\caption{Thumbnail images and 1D spectrum of LAE \#9.
		From left to right: 
		Rc band (Subaru),
		$i'$-band (HST),
		$z'$-band (HST),
		and the spectrum from DEIMOS (Keck).
		The error spectrum is the blue line.}
	\label{f:lae9}
\end{figure*}

The first attempt to search for galaxies
at the redshift of the \civ\ systems
is the spectroscopic study
of \citet{diaz2011}, which followed up  
three galaxy candidates
from \citet{stiavelli2005},
referred to as Target 1, 2 and 3.
Target 1 (J103024.08+052420.41)
was confirmed by its \Lya\ emission
in \citet{stiavelli2005} and 
then by \citet{diaz2011}
at $z_{em}\sim5.973$.
We recalculated the redshift using
our definition based on the bluest pixel
of the \Lya\ line profile and found 
$z_{\text{BLUE}}=5.968$ (Figure \ref{f:lae9}).
This LAE is at 57.163\arcsec\ (335 pkpc) to the South-West 
of the QSO sight-line, thus it is
outside MUSE's FoV
but it lies within 500 \kms\ of the two \civ\ systems
at $z_{\text{\civ}}=5.9757$ and 5.9784,
for which no other galaxy 
candidate was identified in the datacube.
Following the identification number 
assigned to the LAEs in the sample, 
in this work we will refer to this 
galaxy as LAE \#9
(see Table \ref{t:lae9} for details).

\begin{table*}
	\caption{LAE outside the FoV of MUSE.
		Columns are: (1) reference number,
		(2) right ascension hh:mm:ss.ss, 
		(3) declination $\pm$dd:mm:ss.s,
		(4) redshift assuming the blue edge of the line profile,
		(5) angular distance to the QSO,
		(6) total line flux and
		(7) luminosity of \Lya.}
	\label{t:lae9}
	\begin{tabular}{rccccccc}
		\hline
		\multicolumn{1}{c}{ID} &
		\multicolumn{1}{c}{RA} &
		\multicolumn{1}{c}{Dec} &
		\multicolumn{1}{c}{z$_{\text{BLUE}}$} &
		\multicolumn{1}{c}{$\delta\theta$} &
		\multicolumn{1}{c}{$\rho$} &
		\multicolumn{1}{c}{F$_{\text{\Lya}}$} &
		\multicolumn{1}{c}{L$_{\text{\Lya}}$} \\
		\multicolumn{1}{c}{} &
		\multicolumn{1}{c}{(J2000)} &
		\multicolumn{1}{c}{(J2000)} &
		\multicolumn{1}{c}{(redshift)} &
		\multicolumn{1}{c}{(\arcsec)} &
		\multicolumn{1}{c}{(pkpc)} &
		\multicolumn{1}{c}{($\times 10^{-18}$ erg s$^{-1}$ \cm)} &
		\multicolumn{1}{c}{($\times 10^{41}$ erg s$^{-1}$)} \\
		\hline  
		LAE \#9 &   10:30:24.08 & +05:24:20.41 & 5.968 &  57.163  & 335  &  9.3$\,\pm\,$0.8& 38.3$\,\pm\,$3.4\\
		\hline\end{tabular}
	\\
\end{table*}

Target 2 (J103027.98+052459.51, $z_{em}=5.676$) and 
Target 3 (J103026.49+052505.14, $z_{em}=5.719$)
are both inside the MUSE's FoV
but neither were recovered by our detection procedure
based on \Lya\ line emission.
In Appendix \ref{app:im_others},
Figures \ref{f:target2} and \ref{f:target3}
present images of each object
from different observing campaigns.
The top row of images correspond to MUSE
data. The Blue and the Red images are 
broad-band images that sample 
the rest-frame wavelength range
bluer and redder than \Lya,
respectively, with a bandwidth 
defined in 1000\,\kms.
The object cannot be identified in these images
and no emission line is detected in the 2D
spectrum despite the flux value reported by \citet{diaz2011},
F$_{\text{\Lya}}=4.9\pm1.1\times 10^{-18}$, is
above our detection limit.

The second row in Figure \ref{f:target2} 
shows SuprimeCam images in the photometric 
bands R$_c$, NB(\civ), $i'$ and $z'$ from
\citet{diaz2014}. The third row shows
ACS HST images \citep{stiavelli2005}
in i'-band (F775W) and z'-band (F850LP).
As reported in \citet{diaz2014},
Target 2 is detected in $z'$ with S/N$\,\geq 5$
and the signal in the bluer images is too low
to confirm a detection, resulting in red ($i'-z'$) colors and 
supporting the high-$z$ 
nature of the object. However, 
the lack of \Lya\ emission line contradicts
the original classification of this object as an LAE.
Therefore, we will not include this candidate in our 
current analysis.

Figure \ref{f:target3} presents Target 3
which has already been ruled out as high-$z$
source by \citet{diaz2014} based on the detection of the source
in the R$_c$-band with S/N$\sim4$ and magnitude R$_c=26.6\pm0.3$.
Here we review the SuprimeCam images in the second row
of Figure \ref{f:target3} were the source can be identify 
in the R$_c$ thumbnail.
The object is hardly {\bf detected} in the blue and red
MUSE images, and there is no evidence of
an emission line in the 2D spectrum.
This confirms that Target 3 is not at $z_{em}=5.719$ 
and it will not be considered in the future.

In summary, we confirm a lack of \Lya\ emission line
at the flux level reported in \citet{diaz2011} for their
Target 2 and 3, the latter previously ruled-out by \citet{diaz2014}.
Also, we include their Target 1 in the current analysis as LAE \#9.

\subsubsection{\citet{diaz2014,diaz2015}}\label{s:diaz15}

The projected distribution of LBGs brighter than $z'= 25.5$ ABmag 
in the field J1030+0524
revealed that the candidate closest to the QSO line of sight 
is at $\sim5.1$ cMpc ($\sim761 $pkpc).
The absence of bright LBG candidates in the
proximity of the \civ\ system suggests that 
fainter galaxies had to be present at closer distances from
the strong \civ\ system at $z\sim5.7$ \citep{diaz2014}. 
In addition, \citet{diaz2015} confirmed the closest NB-selected
LAE brighter than NB$\sim 25.5$ ABmag at $\sim213$ pkpc
from the \civ\ system, but such a distance
is also too large to account for metal enrichment
given the short age of the Universe at $z\sim5.7$.
As a result, the expectation was that 
the closest galaxies remained below the detection limit.

The new galaxies identified with MUSE and reported in this
work have no continuum detected.
Thus, is in agreement with the large-scale
photometric study of \citet{diaz2014},
we confirm the absence of UV-bright
galaxies within $\sim 250$ pkpc
not only at $z\sim5.7$ but also for all the other \civ\
systems in the line of sight.
Finally, the current sample is a confirmation
of the presence of several very faint galaxies 
($L_{\text{ \Lya}} \sim 0.9 - 5.2 \times 10^{42}$ ergs s$^{-1}$, see Section \ref{s:lumfunc})
within 200 pkpc of \civ\ systems at $z>5.5$,
which agrees with the conclusion from \cite{diaz2015}
at $z\sim5.7$.

\subsubsection{ \citet{cai2017}}\label{s:cai17}

Based on HST imaging, 
\citet{cai2017} reported two narrow-band selected LAE candidates
for the \civ\ systems at $\zciv=4.948$ and 5.744.
The estimates of the \Lya\ luminosity
of these candidates are within the detection limit of our sample
and they should be detectable in the data cube.
We reviewed the data at the coordinates reported
in \citet{cai2017} and found that
the first candidate (Object 1, RA=10:30:26.746 Dec.=+5:24:59.76),
which was associated with the \civ\ system at $z_{\civ}=5.744$,
is 3.1\arcsec away from LAE \#3 (see Figure \ref{f:cai1})
at $z_{\text BLUE}=5.518$.
The left image in Figure \ref{f:cai1} 
is a NB(\Lya) image at the wavelength of \Lya\ 
at z=5.744. There is no evidence of an emission line
at the coordinates for Object 1 in this image.
We also analyzed the white image integrated over
the full wavelength range, the 1D spectrum and the 2D spectrum
and we find no evidence of an emission line.
However, candidate LAE \#3 in our sample is
3.1\arcsec\ from the position of Object 1.
The center panel of Figure \ref{f:cai1}
shows a NB(\Lya) image at the redshift of LAE \#3. 
The \Lya\ emission of LAE \#3 is
detected at $\lambda\sim7924$\,\AA\
which is in the blue wing of the 
NB filter F853N used in \citet{cai2017}.
Therefore, it is possible that we have 
identified the same object, in which case
the first LAE candidate in
\citet{cai2017} would correspond to LAE \#3
in our sample (Table \ref{t:candidates}),
and its true redshift is $z_{\text {BLUE}}=5.518$.

The second object (RA$=10:30:26.746$, Dec.$=+05:24:59.76$)
was identified in the data cube, although it
shows a small offset in our world coordinate system.
Figure \ref{f:cai2} presents a close look at this galaxy.
The top left image, where the object is detected near the center of the
field, was obtained by combining all frames
at wavelengths bluer than the rest-frame Lyman limit
($ \lambda <912$ \AA) at $z_{\text{\civ}}=4.948$. 
The top center shows the NB image for \Lya\ at 
 $z_{em}=4.948$ and there is no emission line
 on it. The 2D spectrum in the middle panel
 shows a faint continuum
 signal across the wavelengths range
 that would correspond to \Lya\ at $z_{em}=4.948$.
In this case, we can rule out the high-$z$ LAE hypothesis
based on the detection of
flux at rest frame $\lambda < 1216$\AA .

\subsection{Environment of absorption systems}\label{s:pairs}

\begin{figure*}
	\includegraphics[scale=0.5]{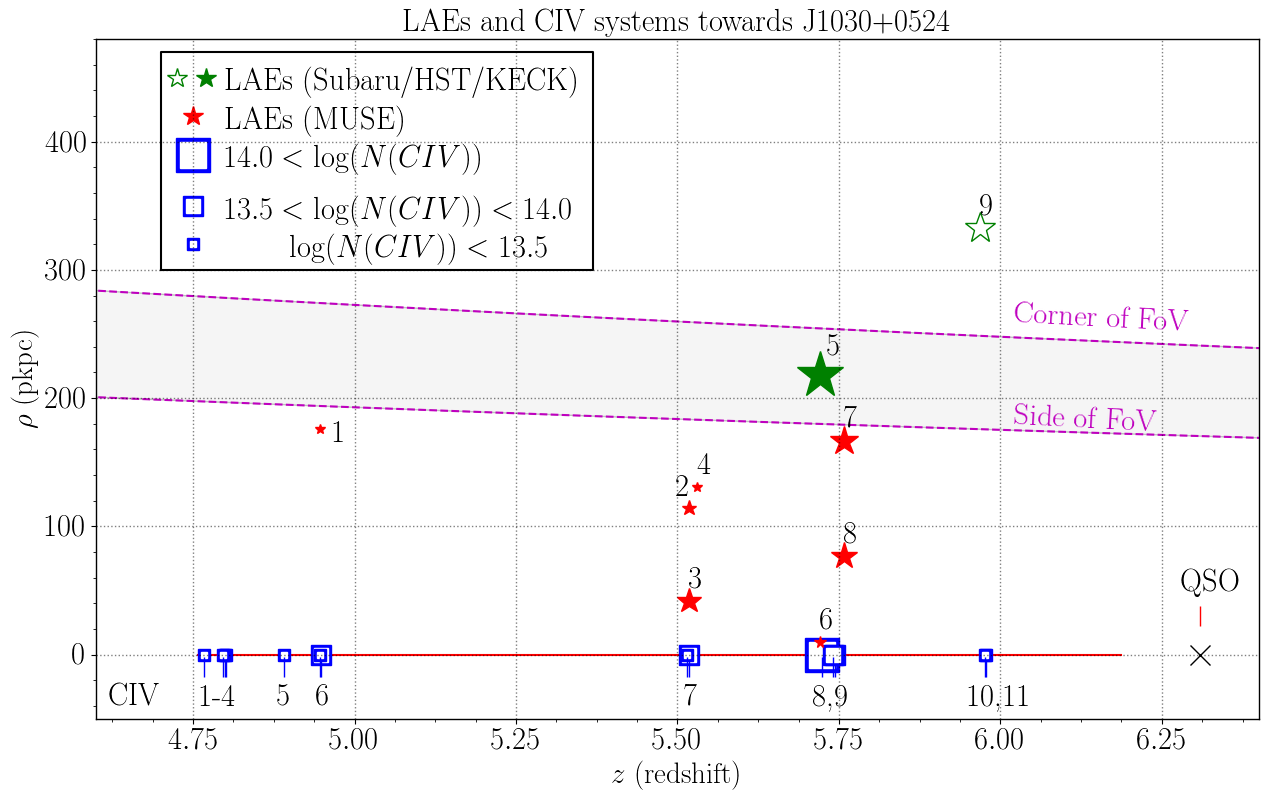}
	\caption{Representation of the spatial distribution of
		the LAEs respect to the \civ\ systems in J1030.
		The $y$-axis is the transverse distance in proper
		kpc and the $x$-axis is redshift.
		The QSO's redshift is indicated with an ``X'' and 
		the absorption path is indicated with the red solid line at $y=0$.
		Squares' sizes indicate the column density of the \civ\ absorption
		and the number is the ID from \citet{dodorico2013}.
		Green star symbols are LAEs from our 
		previous work, red star symbols are 
		newly discovered LAEs.
		The size of the stars indicates the \Lya\ flux
		(see Tables \ref{t:candidates} and \ref{t:lae9}).
		The FoV of MUSE (purple dashed line) 
		shows the limit in transverse distance
		to search for \civ-galaxy associations
		imposed by the size of the FoV.}
	\label{f:qso-los}
\end{figure*}

The X-Shooter spectrum of the QSO J1030+0524
revealed 11 \civ\ systems in the range
$z_{\text{\civ}}=4.757$--6.187 \citep{dodorico2013},
which are listed in Table \ref{t:civ}.
In the sample, there are three systems 
with double \civ\ absorptions
(\civ\ ID: 6, 7 and 9), one strong system 
( \civ\ 8, $\log_{10}(N_{\text{\civ}}[$\cm$])>14.0$),
and seven weaker
\civ\ systems (\civ\ 1, 2, 3, 4, 5, 10 and 11, $\log_{10}(N_{\text{\civ}}[$\cm$])<13.5$). 
The search for LAEs in the environment of these absorptions
resulted in at least
one LAE for each \civ\ system with
$\log_{10}(N_{\text{\civ}}[$\cm$])>13.5$
and multiple LAEs for the systems at $z_{\text{\civ}}>5.5$.

One of the main results is summarized in
Figure \ref{f:qso-los}, it shows the impact parameter
of the LAEs with respect to the line of sight
as a function of redshift.
The \civ\ systems are represented by open squares
of sizes divided into three bins in column density.
The LAEs are represented by star symbols with sizes
proportional to the flux of the \Lya\ emission,
and labels indicating the ID numbers.
Green symbols correspond to galaxies previously known 
and red symbols are new detections. 
The dashed lines indicate the radial distance
from the position of the QSO in the center of 
MUSE's FoV to the edge and the corner 
of the FoV.
This shows that
MUSE can cover about $\sim200$-280 pkpc
and within this distance we have found multiple
LAEs for three \civ\ systems at $z> 5.5$.
Also, all \civ\ 
systems with $\log_{10}(N_{\text{\civ}}[$\cm$])>13.5$
at least one LAE candidate 
within $\sim200$-280 pkpc.

\begin{figure}
	\includegraphics[scale=0.45]{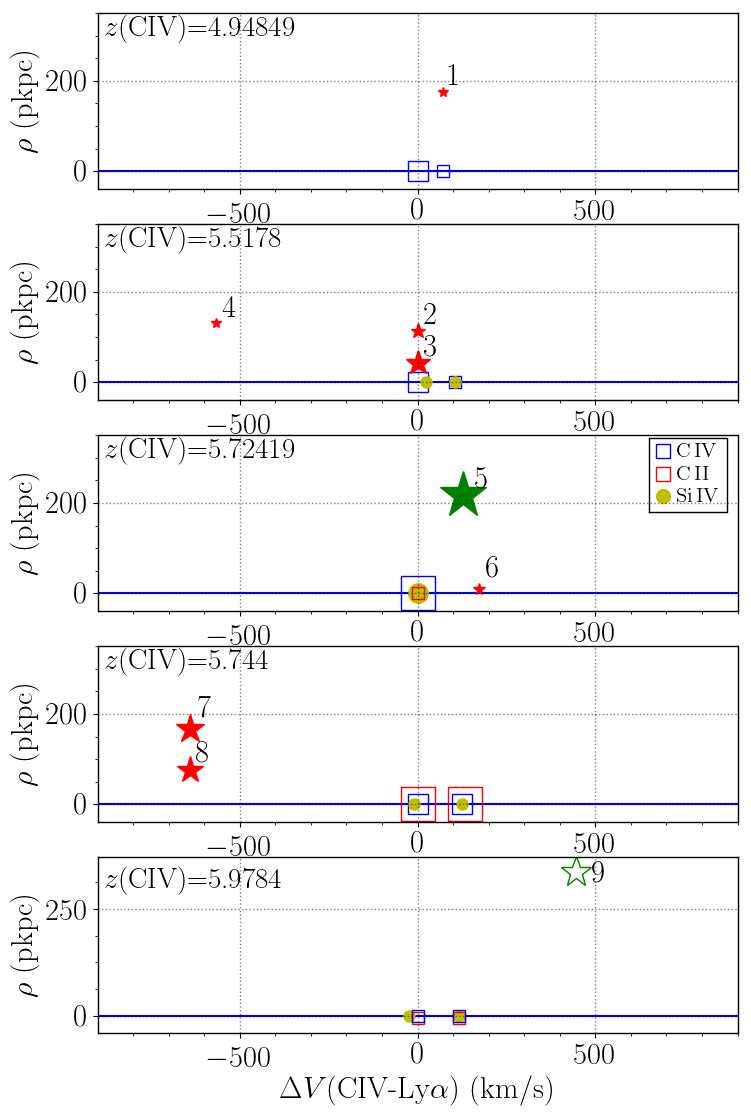}
	\caption{Impact parameter vs. radial velocity 
	    difference with the nearest \civ\ system.
		Three \civ\ systems have LAEs within $\pm 250$\,\kms.
		Note that all \Lya\ redshifts are measured from the 
		blue edge of the \Lya\ profile, thus the systemtic redshift of each galaxy is likely to be slightly 
		lower than reported, resulting in positive (negative) $\Delta$V values likely being higher (lower)
		than reported. In the case of the LAEs at 
		$\approx -600$\,\kms, the velocity difference 
		would reduce. Using the empirical calibration 
		developed by \citet{verhamme2018}, and typical 
		FWHM velocities of 200 \kms, we expect each 
		$\Delta$V value to be offset by 140 \kms\ 
		to the right in the diagram above. Thus, 
		all \civ\ systems have an LAE with a 
		corrected redshift with 500 \kms.}
	\label{f:vel_rho}
\end{figure}

Providing a closer look,
Figure \ref{f:vel_rho} shows the 
line of sight velocity to the corresponding \civ\
vs. impact parameter. 
In each case, the redshift of the \civ\
is indicated in the top left corner.
The \civ\ systems are represented by open blue squares 
at $\Delta V(\text{\civ - Ly}\alpha)=0$
that scale with column density as in Figure \ref{f:qso-los}.
The presence of \cii\ and \siiv\ is indicated with
open red squares and filled circles, respectively.
In three cases, the closest LAEs
are also within $\pm250$\,\kms\ of
the corresponding \civ. 
In the first case, \civ\,6 is a
double system with one LAE
at 176 pkpc and almost the same redshift 
($\Delta V(\text{\civ - Ly}\alpha)=71$\,\kms) 
The distance is not large enough to rule out  
a galactic wind scenario.
Stepping forward in redshift, the double system \civ\,7 at $z_{\text{\civ}}=5.5172$
has three associated LAEs: 
LAE \#2, \#3 and \#4. The first two lie at less than $\sim2$ \kms\
from the absorption (second panel of Figure \ref{f:vel_rho}).
In particular, the closest galaxy LAE \#3 at 42 pkpc is also
the brightest \Lya\ emission of the three. 
Moreover, the absorption system presents weak
\siiv\ in both of the components.
Similarly, \civ\,8 at $z_{\text{\civ}}=5.72419$ 
has two associated LAEs at less than 172 \kms: LAE \#5,
which was previously reported in \citet{diaz2015}
and remains the brightest galaxy in the LAE sample of this work,
and LAE \#6, which is a new detection 
at $\rho=10$ pkpc to the \civ\ system.
Figure \ref{f:pos-all} shows that 
LAE \#5 and \#6 are almost at opposite sides of the QSO's line-of-sight
separated by $\sim 223$ pkpc in projected distance,
and $\sim 50$\,\kms\ in velocity.
Thus it is not unreasonable to assume that these 
two galaxies are gravitationally bound, 
and that LAE \#6 being the fainter 
of the two is a minor companion or 
satellite of LAE \#5 (see section \ref{s:IGM} for further discussion).

For the double system \civ\,9 at $z_{\text{\civ}}=5.7428$ 
there are two confirmed LAEs at
167 pkpc (LAE \# 7) and at
77 pkpc (LAE \#8), both at $-639$\,\kms .
The asymmetry of the emission lines suggests
that the true line centroids are bluer than our measurement 
thus the module of the velocity difference is only a lower limit in this case 
(the LAEs are likely closer).
It is interesting to note that this particular system, 
with the largest velocity difference to the LAEs, is the only system
with strong \cii\ absorption (even higher column density than 
the \civ\ component). 

Finally, the bottom panels of Figure \ref{f:vel_rho} present the position 
of LAE 10 with respect to \civ\ 10 and 11. 
This LAE is clearly too far from the \civ\ to be the source of metals
observed in absorption. However, the fact that two weak \civ\ systems with
some evidence of \siiv\ have been found at less than $\sim 530$\,\kms 
is an indication that such \civ\ systems are associated with the environment of 
LAE 10, although the true sources of these absorptions are still to be found.

\subsection{\civ\ equivalent width and LAE impact parameter}\label{s:ew-d}

\begin{figure}
	\includegraphics[scale=0.45]{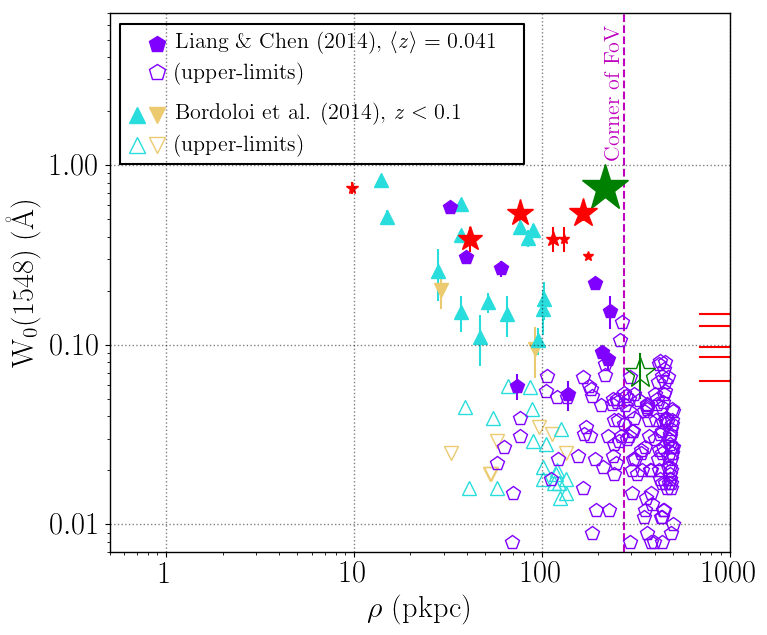}
	\caption{Rest-frame equivalent width 
		of \civ\ versus impact parameter (transverse distance). 
		Comparison with low mass (dwarf)
		local ($z< 0.1$) galaxies 
		from \citet{liang2014} (open and solid pentagons)
		and \citet{bordoloi2014} (open and solid triangles).
		The vertical dashed line indicates 
		the maximum distance from the center of the field of view 
		(QSO's line of sight) to the corner.
		The horizontal line markers on the right side of each panel
		indicate the column density of 
		four \civ\ systems from \citet{dodorico2013} for which 
		we do not find a galaxy counterpart.
		The open circles indicate if the LAE is the closest detection
		to the corresponding \civ.}
	\label{f:ew-dist}
\end{figure}

Clear evidence of the existence of a chemically enriched
CGM around star forming galaxies comes from
the observed relation between W$_0$ of absorption lines of 
different metals and the impact parameter ($\rho$) to their 
closest neighbor galaxy 
\citep{adelberger2005b,steidel2010,bouche2012b}.
Figure \ref{f:ew-dist} presents this relation for 
all the LAEs in this work (red star symbols)
and provides a comparison
with observations at $z<0.1$.

We have found LAE counterparts of four of the five
\civ\ systems with W$_0(\text{\civ})> 0.2$ \AA,
within $\rho\sim270$ kpc (Figure \ref{f:ew-dist}).
Regarding the six \civ\ systems with W$_0(\text{\civ})< 0.2$ \AA,
none of them have a galaxy detection in MUSE's data-cube
(horizontal red lines to the right of Figure \ref{f:ew-dist}),
although two of them are neighbors of LAE \#9 which is outside MUSE's FoV
at $\sim 335$ pkpc.
Although the sample is small and affected by low number statistics,
it seems that systems with W$_0(\text{\civ})> 0.2$ \AA\
at $z=5$--6 are likely to have galaxies with \Lya\ emission
within $\rho< 200$ pkpc (4/5 cases),
whereas the \civ\ systems with W$_0(\text{\civ})<0.2$ \AA\
do not seem to have a galaxy with detectable \Lya\ emission (0/6 cases). 
Therefore, 
{\it it is likely that the true sources
	of these weak absorption systems remain below the detection limit.}

\subsection{\Lya\ luminosity distribution}\label{s:lumfunc}

\begin{figure}
	\includegraphics[scale=0.36]{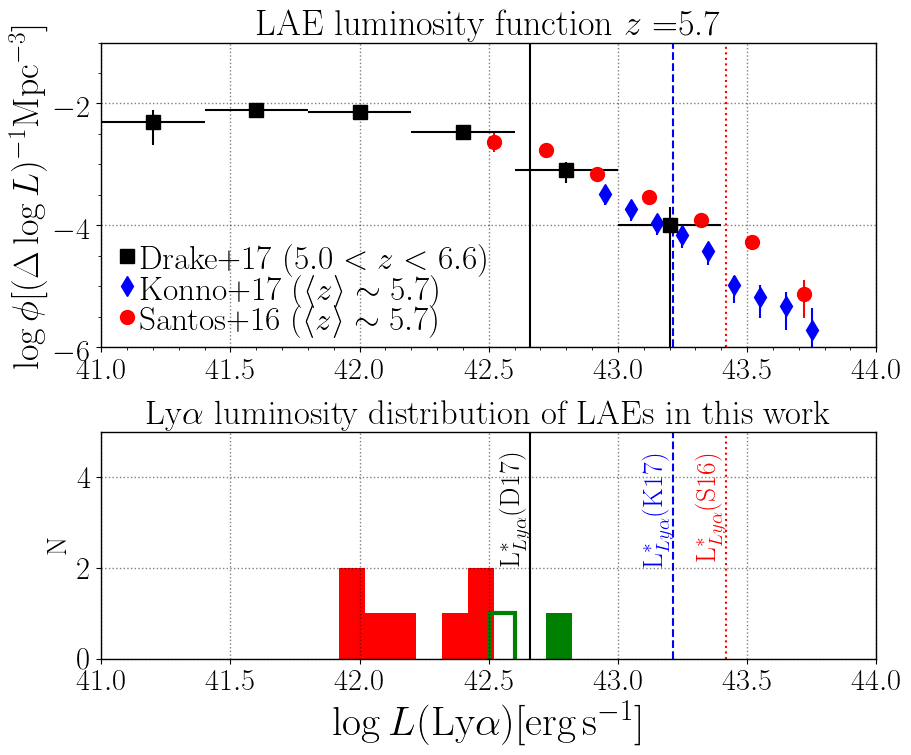}
	\caption{{\bf Top:} Recent estimations
		of the luminosity 
		function of LAEs at $z=5.7$
		from \citet[][blue squares]{konno2017}
		and \citet[][red circles]{santos2016}.
		{\bf Bottom:} The distribution of the 
		\Lya\ luminosity of LAEs in this work.
		The solid green bar represents
		LAE \#5, the open green bar represents LAE \#9
		and the red histogram corresponds to the newly reported
		LAEs from MUSE's data,
		showing that these type of galaxies 
		are the population in the faint end of the most recent
		luminosity functions displayed in the top panel.}
	\label{f:lum-func}
\end{figure}

One of the most interesting results is the low \Lya\ luminosity
of the candidates. Extensive efforts have been made
in the past to study high-redshift galaxies in this field
\citep{stiavelli2005, kim2009, morselli2014, diaz2014, diaz2015,cai2017}.
However, none of the previous studies reported the galaxies 
we detected with MUSE, except for LAE \#5.
The reason is that our LAEs are below the detection limits of recent works
based on narrow-band and broad-band photometry
\citep[e.g.][]{santos2016, konno2017}.
The top panel Figure \ref{f:lum-func} 
shows the luminosity function of $z=5.7$ LAEs
from \citet{santos2016}, \citet{konno2017} and \citet{drake2017}.
The bottom plot presents 
the luminosity distribution of the LAEs from this work.
Solid red histogram is for new detections with MUSE data,
and green is for LAEs \#5 and \#9.
We find that the LAEs
in the environment of \civ\ systems trace 
the extrapolation of the faint end slope of 
the \Lya\ luminosity function,
with some galaxies 0.5 dex fainter than 
the large LAE samples
used by \citet{santos2016} and \citet{konno2017}.
In comparison with NB-selected samples, 
our candidates are in the range of
L$_{{\text \Lya}}= 0.03$--$0.32$ L$^{\star}_{{\text \Lya}}$,
with $\log($L$^{\star}_{{\text \Lya}})= 43.42$ \citep{santos2016}
and $\log($L$^{\star}_{{\text \Lya}})= 43.21$ \citep{konno2017}.

The \Lya\ luminosity function measured with MUSE
in the HUDF has probed the existence of
LAEs that are even fainter than our sample, which are
accessible only to wide-field IFUs.
\citep{drake2017} reported 
 $\log(L^{\star}_{{\text \Lya}})= 42.66$  based on 
 137 hours of MUSE observations
covering 3\arcmin$\times$3\arcmin area.
According to this new estimation,
the luminosity of our sample is in the range
L$_{{\text \Lya}}= 0.18$--$1.15$ L$^{\star}_{{\text \Lya}}$,
of which the only galaxy brighter than L$^{\star}_{{\text \Lya}}$
is LAE\# 5 (solid green bar in Figure \ref{f:lum-func}).
In summary, the new sample of LAEs obtained with MUSE in the field
of QSO J1030+0524 is fainter than L$^{\star}_{{\text \Lya}}$.

\begin{figure}
	\includegraphics[scale=0.4]{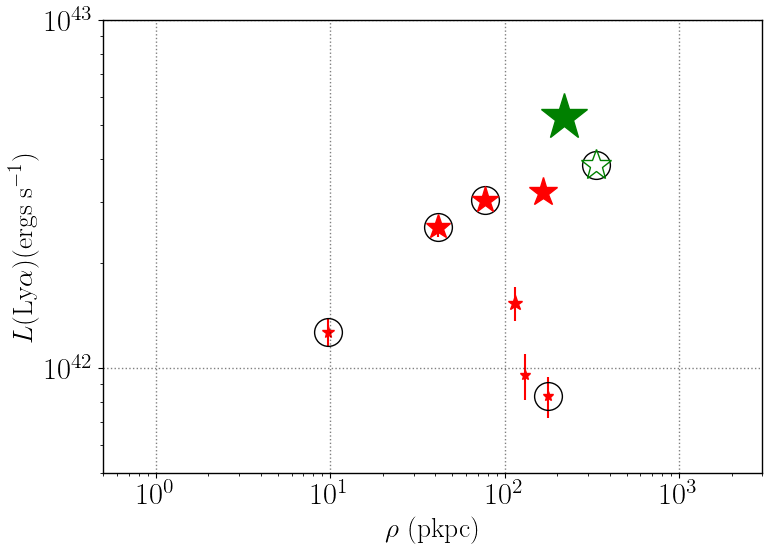}
	\caption{Luminosity of the \Lya\ emission as a function 
		of impact parameter. The circles indicate the closest 
		LAEs to the corresponding \civ\ system.}
	\label{f:lum-dist}
\end{figure}

Finally, there is tentative evidence for a trend of fainter
LAEs at smaller impact parameters.
This is presented in the \Lya\ luminosity vs. impact parameter
plot of Figure \ref{f:lum-dist}, which shows that LAEs at larger 
impact parameters are also more \Lya\ luminous. 
The circles indicate the LAEs with the smallest 
impact parameter for a given \civ\ system.

\section{Discussion}\label{s:discussion}

\subsection{Faint optical galaxies polluting the IGM}\label{s:IGM}

The brightest LAE in the sample is LAE \#5 
(L$_{\text{\Lya}}= 1.15 $\,L$^{\star}_{\text{\Lya}}$),
previously found at the same redshift 
of the \civ\ with the highest W$_0$ of the sample.
However, the impact
parameter $\rho\sim218$ pkpc is in 
conflict with the outflow scenario
for the origin of the absorption
because, assuming formation at $z=30$,
a galaxy that has been forming stars for 
less than $\sim900$ Myr would require high speed 
outflows to enrich such distances in a relatively
short time  \citep{diaz2015}.
The alternative scenario implies that the origin
of the absorbing gas is a fainter (previously undetected) 
galaxy at a closer distance from the \civ\ system.
\citet{garcia2017a} also concluded that the ``dwarf 
satellite outflow'' scenario is favored by 
state-of-the-art hydrodynamical simulations
of the epoch of reionization.
In this context, the detection of LAE \#6
with L$_{\text{\Lya}}=0.28$\, L$^{\star}_{\text{\Lya}}$
at $\rho=10$ pkpc from the \civ, 
provides evidence for CGM-IGM 
chemical enrichment by satellites
and neighbor dwarf galaxies at very early times,
as predicted by computer simulations
\citep[e.g. ][]{madau2001,shen2012,garcia2017a}.

The LAEs with smaller impact 
parameters to the \civ\ systems at $z>4.9$ 
are good candidates for the source of the metals
because metal-enriched winds 
departing from the source at $z=10$
could have traversed this distance
without the need for extreme outflow velocities.
Figure \ref{f:ew-dist-wind} shows
in blue dotted lines the distance 
that an outflow of mean speed 
$\langle V_{wind} \rangle=100$, 200, 300 and 400\,\kms\
would cover if starting at $z=10$.
Except for LAE \#1,
the closest galaxies to the \civ\ systems
are below the 200\,\kms\ dotted line,
which makes them good candidates 
for the sources of the metals.

The absence of LAE candidates
within $\rho=200$--250 pkpc
(i.e. inside MUSE's FoV) 
of five \civ\ systems with 
W$_0(\civ)<0.2$ \AA\ is difficult to 
be explained by strong outflows from galaxies
at $\rho>250$ pkpc (i.e. outside MUSE's FoV).
Given the short age of the Universe at these redshift,
winds of several hundreds \kms\ must be active
over most of the galaxy formation
history to deliver metals at distances
$\rho>250$ pkpc.
For example, LAE \#9 
at $\rho \sim 335$ pkpc 
is likely not the source of metals
for \civ\ 10 and 11.
One interpretation of this result
is that there are galaxies 
below the detection limit 
of this work ($ 2\times 10^{-18}$ erg s$^{-1}$ \cm)
which are probably close to the absorption systems.

It is possible that what we see at $z>4.75$ is the 
early contribution of low luminosity and low mass
galaxies to the metal content of the CGM-IGM.
Since mass estimates for the high-$z$ LAEs
are not available, based on the mass-luminosity relation
$M_{\text{STAR}}/M_{\odot}$--M$_{\text{UV}}$
for $z=4$--8 UV-selected galaxies from \citet{song2016} 
and the detection limit of M$_{\text{UV}}<-20.5$ mag 
from \citet{diaz2014}, we estimate a conservative upper limit
of $\log(M_{\text{STAR}}/M_{\odot})\simlt 9.5$
for the newly reported LAEs in our sample 
(i.e. excluding LAE\# 5 and LAE\# 9). 
For example, the range of \Lya\ luminosities of our LAEs
 L$_{\text{\Lya}}= 0.18$--0.84 \,L$^{\star}_{\text{\Lya}}$
has been associated to $z>3$ galaxies in the range 
$\log(M_{\text{STAR}}/M_{\odot})\simlt 8.5$ \citep[][]{karman2017}.
Although the physical properties of low-mass high-$z$ 
galaxies are under debate, several authors agree 
that these faint LAEs have typically very young ages, 
high specific star-formation rates, blue UV-continuum 
slope, low metallicities and low dust extinction 
\citep[e.g.][]{trainor2016,karman2017,amorin2017,debarros2017}.
Therefore, \civ\ systems at these redshifts could
be tracing gas in the proximity of young and small galaxies,
similar to the currently best candidates
to drive the epoch of reionization at $z>6$ 
\citep[][]{robertson2015,atek2015,castellano2016,livermore2017}.

\subsection{Connection with UV background}\label{s:UVB}

The UV ionizing background radiation field (or UVB)
became homogeneous at the Lyman limit 
rapidly after $z\sim5.5$. 
During reionization and probably down to $z\sim5$,
the UVB was not spacially homogeneous 
\citep{fan2006, becker2015a, bosman2018, eilers2018}. 
Large spatial fluctuations are predicted for
the intensity and slope of this radiation,
on scales of tens of comoving Mpc 
\citep{mesinger2009, davies2016, chardin2017, daloisio2017, keating2018, uptonsanderbeck2019}. 
Indeed, the discovery of one 
particularly long and opaque 
\lya\ trough \citep{becker2015a} 
led to observations of a derth 
of $z\sim5.7$ LAEs within 20 h$^{-1}$ 
comoving Mpc, confirming the prediction 
that the scatter in \lya\ opacity 
is caused by spatical variation 
in the UVB over large scales \citep{becker2018}. 
The \lya\ opacity towards J1030+0524 
has been measured by \cite{eilers2018} 
at z=5.74 to have an average flux of 
$<F>=0.0144$, which corresponds 
to a mean optical depth of $\tau_{\rm eff}=4.2$. 
We can use the mean surface density of
LAEs in the field J1030+0524 within 
10 h$^{-1}$ cMpc of 0.107$\pm$0.004 per square arcmin 
\citep{diaz2014} (noting this is a lower limit 
at a sensitivity of m$_{\rm NB}(5\sigma$)=25.6) 
to compare directly with Figure 14 
in \cite{davies2018}. This middle-of-the-road 
\Lya\ opacity -- \cite{eilers2018} measures 
an average $\tau_{\rm eff}=4.0057\pm 0.2469$ at $z=5.75$) --  
has less leverage to test models which 
predict an associated under or over-density 
of galaxies. The measured opacity and galaxy 
density along this line of sight is consistent 
with both the fluctuating UVB \citep{davies2018} 
and fluctuating temperature models \citep{daloisio2017}.

Such observations of \lya\ opacity reflect 
the state of the UVB at 13.6 eV, 
whereas the energy required to ionize 
\ciii\ to \civ\ is 47.89 eV. Ideally,
to probe the UVB at this energy, 
\civ\ itself should be used, 
this is somewhat complicated by the fact 
that \civ\ traces the metals around galaxies.
\cite{meyer2019} cross correlated 
37 \civ\ systems at $4.3<z<6.2$ with 
\Lya\ forest flux and found that the UVB 
is enhanced on scales greater than 10 cMpc 
(as measured by a decrease in forest absorption),
whereas an excess of \lya\ absoprtion is seen 
in the immediate vicinity of \civ\ systems. 
Furthermore most of the photons that ionize 
\ciii\ to \civ\ also have the energy to ionize \HeII.
However, the Universe is optically-thin 
to \HeII\ only after $z\sim3$. 
Therefore, the UVB relevant for \civ\ 
is highly inhomogeneous at $z\sim5.7$
despite \HI\ reionization completing around 
$z=5-6$. As a result, in the presence of 
an evolving ionizing background modulated 
by the environment, the detection of extended 
ionized metal halos around galaxies would
also show a dependency with the ionzing
properties of the local UVB.

The line of sight to J1030+0524
intercepts an excess of narrow-band selected 
LAEs at $z=5.7$, and a deficit of broad-band 
selected LBGs, on scales of 10 cMpc 
with respect to the mean number density of galaxies on 
a $80\times60$ cMpc FoV \citep{diaz2014}.
This was interpreted as an indication that 
an increase in \civ\ number denisty could
be associated with a medium-density environment 
dominated by young star-forming galaxies 
on large scales. 
The current work presents evidence 
in support of the idea that strong \civ\ 
at $z>5.5$ is biased 
to highly ionized regions
for the following reasons.
We have found that the three systems at $z>5.5$
with W$_0(\civ)>0.5$\,\AA\
and double \civ\ and \siiv\ components,
have two or three LAEs within 200--220 pkpc.
In addition, these galaxies are in the range
L$(\Lya)= 0.18$--1.15 L$^{\star}_{\text{\Lya}}$,
mostly tracing the faint end of the \Lya\ 
luminosity function (Figure \ref{f:lum-func}).
These kind of LAEs have been found to have 
narrow line widths associated to
low \HI\ column densities \citep[e.g.][]{guaita2017},
strong optical emission lines associated to high
ionization parameters \citep{nakajima2014},
extremely high nebular excitation \citep{trainor2016},
and very high sSFR \citep{karman2017}.
Therefore, the narrow \Lya\ emission reveals the existence of 
recent star-formation under conditions that could
favor the escape of ionizing 
radiation from the galaxy.
Moreover, the thorough 
inspection of narrow band images from stacked frames
described in section \ref{s:detection} did not returned any
UV bright galaxy in MUSE's FoV. 
This result supports previous conjectures that the 
\civ\ systems at the highest redshifts could be 
tracing a region of the Universe
where large UV bright galaxies have not yet formed.

The surroundings of bright galaxies 
are guaranteed to be more enriched than the environment
of faint galaxies beacause faint low-mass galaxies cluster around 
bright galaxies, and metallicity is predicted to
increase monotonically with gas overdensity \citep[e.g.][]{keating2016}.
As a consequence, the prevalence of \civ\ near faint LAEs
would most likely be related to an overwhelming abundance of faint galaxies 
and/or their contribution to the local UVB rather 
than an generalized deficit of metals around bright galaxies.

In $\lambda CDM$ cosmology, the same faint 
dwarf galaxies that are expected to 
power reionization are also predicted to pollute 
the IGM/CGM with their 
heavy elements, which can be easily ejected outside these galaxies 
because of their shallow potential wells 
\citep[e.g.][]{madau2001,choudhury2008,salvadori2014}. 
Supernova-driven outflows inject metals into 
the CGM/IGM, heat gas that would otherwise fuel 
the next generation of star formation 
(keeping the galaxies faint) and may allow 
for the escape of a large fraction 
of Lyman continuum photons
\citep[e.g.][]{ferrara2013,wise2014}.
A similar scenario has been proposed for the high-$z$ progenitors 
(satellites) of the Local Group \citep[e.g.][]{salvadori2014}, 
which at $z\sim 5$,6 are predicted to be visible as faint LAEs 
\citep[\Lya $= 10^{39-43.25}$ erg s$^{-1}$][]{salvadori2010}.
Therefore, their connection to high-ionization absorption systems
is potentially a consequence of the role of satellites in the early
evolution of the CGM and the IGM.

The potential of faint galaxies boosting the
local UVB at $z\sim5.75$ is suggested by the latest
results from the cosmological hydrodynamic Technicolor Dawn 
simulations. \citep{finlator2020} explore the environment of
strong \civ\ systems at $z > 5 $ and reports that 
a density-bounded escape fraction model
which hardens the emerging flux from galaxies,
results in an ionizing background that produces
the observed volume-averaged hydrogen ionization rate
while boosting the predicted \civ\ 
abundance into better agreement with observations
than previous models.
In addition, they predict that galaxies and strong \civ\ systems 
are positively correlated out to $\sim 300$ proper kpc, 
with $\sim 2$ faint LAEs ($\log($L$_{\text{\Lya}}) > 40.8$)
within 100 pkpc at $z = 5.75$.
Our results are in total agreement with this predictions:
although we do not reach the \Lya\ luminosity limit
of  $\log($L$_{\text{\Lya}}) > 40.8$, we have found 
multiple LAEs within $\sim 250$ pkpc of three
strong \civ\ systems (W$_0>0.2$\AA) at $z>5.5$,
with the faintest examples lying at $\rho <100$ pkpc.

If the common assumption that LyC photons are easier to escape 
from low mass galaxies is confirmed \citep[e.g.][]{bian2017},
low luminosity LAEs would be a very common source 
of ionizing photons at high-$z$.
Therefore, having found several low luminosity LAEs
among the highest redshift \civ,
it is possible that such absorption systems 
are sensitive to the local conditions of 
the ionizing background provided by the LAEs. 
In this case, the highly ionized surroundings 
of these particular galaxies 
would have facilitated the detection of \civ\ in the CGM.
 This opens the posibility that 
 \civ\ systems at $z\,\simgt\,5.5$
could have a preference for recently 
ionized environments,
which should be tested with larger samples
and observations of similar or better depth.

In summary, although the \lya\ opacity 
of this line of sight is average, 
the galaxy environment along this 
\civ-rich line of sight is characterized 
by an enhanced number density of faint LAEs 
near strong \civ, and a lack of 
UV-bright counterparts for 11 \civ\ systems. 
This excess of faint galaxies could boost 
the UVB at energies higher than 1 Rydberg. 
Many more line of sight are required to test 
the connection between \lya\ opacity, 
\civ\ absorbers and galaxies at $z>5.5$.

\subsection{The size of the metal enriched CGM}\label{s:CGM}

The distribution of individual LAEs 
near \civ\ systems at $z=4.95$--5.97
shows a tentative decrease in W$_0(\text{\civ})$ 
with increasing $\rho$
and there is no evidence for a significant
drop at $\rho< 200$.
The accumulation of metals 
in the CGM over time, driven 
by star-formation and outflows 
from the central galaxy 
and its satellites,
leads to a natural growth 
in W$_0$ of the absorption systems
observed near the galaxies.
Therefore, the maximum of the distribution 
of galaxies in the $\rho$ vs. W$_0(\text{\civ})$ 
plane (Figure \ref{f:ew-dist})
would increase with time 
(larger W$_0$ at lower redshift)
and with halo mass, 
which is supported by the 
observations of the mean \civ\ distribution
around bright LBGs at $z\sim2.2$--2.4 \citep{steidel2010}.
However, the evolution with time 
of the maximum detectable distance
(i.e. the size of the \civ\ ``bubble''),
might depend 
on the eficiency of the feedback mechanisms
driving outflows and the energy distribution of
the UVB radiation.

The source of weak \civ\ absorbing clouds
for which no LAE candidate was detected,
could be a \Lya\ emitters below our detections limit, 
or could have stopped emitting \Lya\ photons, 
or the \Lya\ emission is absorbed by intervening 
Lyman Limit Systems
or Damped Lyman Alpha systems (which 
are more common at higher redshifts). 
These alternatives have in common 
that the source would be an 
object in the low-mass regime,
implying outflows with
average low speeds and/or short travel times that
result in a low covering fraction 
of ionized metals.
These would make the \civ\ absorption
detectable only at small impact parameters 
from the source.
Therefore, the missing sources are likely to be close 
(within a few tens of kpc) to the absorbing gas.
These faint galaxies are 
only accessible to current capabilities 
through very deep observations 
\citep[e.g. $>10$ hours with MUSE,][]{drake2017}
or through gravitational lensing
\citep[e.g.][]{karman2017}.
 
However,
studies of low redshift galaxies using background QSOs
have demonstrated the high incidence of \civ\
in the CGM of low luminosity and low mass galaxies.
For example, the 43 galaxies in the COS-Dwarfs program 
revealed that \civ\ is detected out to $\sim 100$ pkpc 
in dwarf galaxies at $z<0.1$ \citep{bordoloi2014}. 
The \civ\ in the CGM of these
faint galaxies could account for $\sim60$\% of the W$_0(1548)>0.1$\AA\ 
absorptions at low-$z$.
\citet{liang2014} study the low- and high-ionization 
metal transitions in the CGM of 195 isolated galaxies 
at $\langle z\rangle =0.041$ over a wide stellar mass range.
They find a large contribution of dwarf galaxies to the \civ\ content
of the CGM, and a general absence of 
heavy elements at $\rho >0.7 R_{\text{h}}$ (where $R_{\text{h}}$
is the dark mater halo radius).
Figure \ref{f:ew-dist} 
shows that the distribution of high-$z$ 
LAEs and \civ\ systems in the 
W$_0$ vs. $\rho$ plane\footnote{Values reported in the literature
only include the equivalent width of 
the stronger line component of 
the \civ\ doublet at 1548 \AA .} 
is in good agreement with 
the low redshift samples of sub-L$^{\star}$ galaxies
at $z\simlt0.1$ from \citet{bordoloi2014}
and \citet{liang2014}.
If \civ\ absorptions
at $z>5$ and $z<0.1$ trace the CGM of 
dwarf galaxies in low-density environments,
it is worth to consider the possibility 
to search for local analogs of reionization epoch galaxies
based on low mass galaxies
with high column density of photoionized 
gas in the CGM.
In order to obtain a truly unbiased view of the 
evolution of the CGM,
the comparison
of galaxies near \civ\ detections 
at different redshifts
requires a blind survey of galaxies 
in fields with detected 
metals absorption systems.

\begin{figure*}
	\includegraphics[scale=0.5]{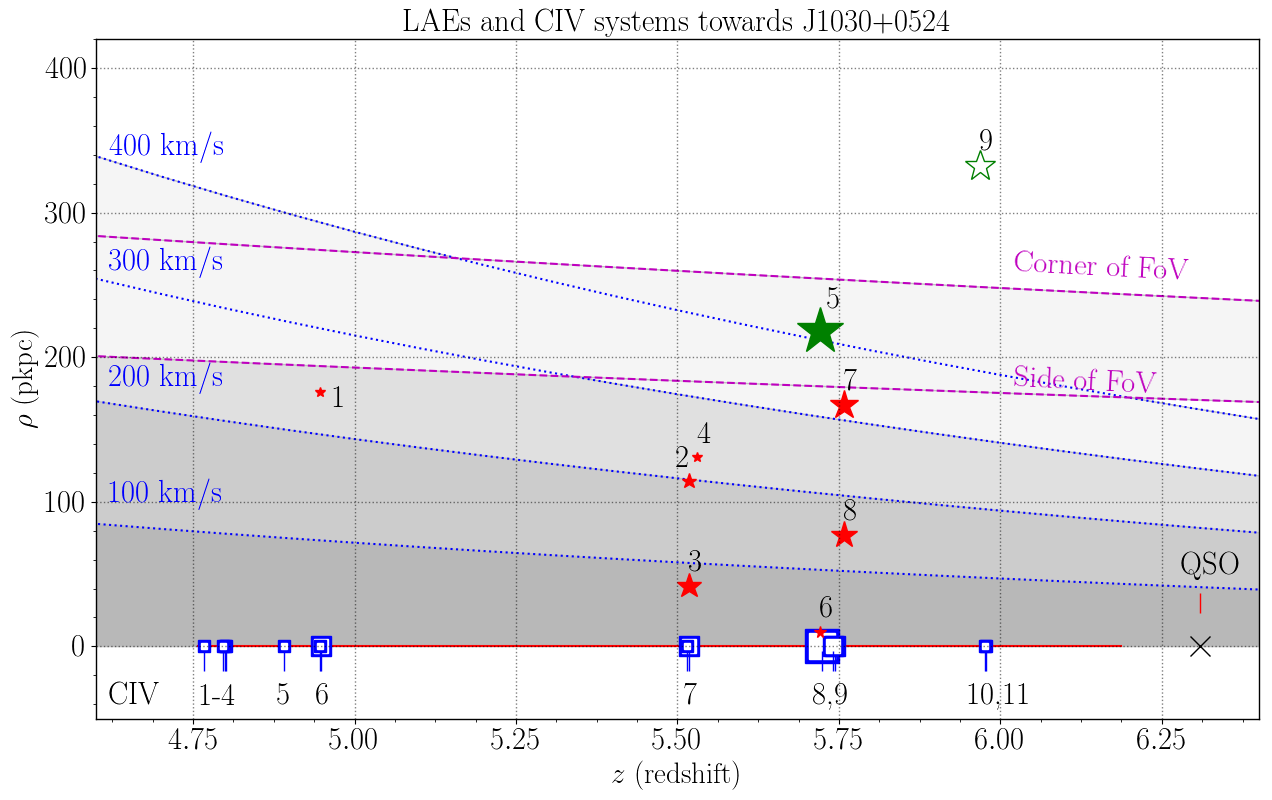}
	\caption{Reproduction of Figure \ref{f:qso-los}
		with shaded areas indicating the distance reached by an outflow starting at $z=10$
		and mean speed $\langle V_{wind}\rangle=100, 200, 300$ and 400\,\kms.}
	\label{f:ew-dist-wind}
\end{figure*}

\subsection{Implications for chemical enrichment}

Our findings that \civ\ systems are in regions with
more than one sub-L$^{\star}_{\text{\Lya}}$ LAE, 
is an indication that \civ\ systems are tracing
environments of recent star-formation.
This result has important implications for the chemical
enrichment of the Universe. 
As we look back in time, the mean metallicity 
of the Universe decreases predictably as a function
of the star formation rate density, moderated by 
the yeild and return fraction of metals \citep{madau2014}. Ideally, our obeservations 
would capture a full census of all ions, 
but in practice they are limited to those 
with favourable transitions. 
The evolution of the comoving 
mass density of \civ, \Ociv, 
reflects both the metallicity 
of intervening IGM and CGM asborbers 
as well as the fraction of Carbon 
in the triply ionzied state. 
The value of \Ociv\ declines mildly 
over the 12.5 billion years from 
redshift $z\sim0$ to $z\sim5$ 
\citep{songaila2001, scannapieco2006, danforth2008, cooksey2010, cooksey2013, dodorico2010, boksenberg2015}. 
In contrast, \Ociv\ appear to decline considerably
over a much shorter period of time, 300 million
years, from redshift 5 to 6 
\citep{ryan-weber2009, becker2009, simcoe2011, dodorico2013, diaz2016, codoreanu2018, meyer2019}.
Despite considerable observational effort to
determine this drop in \Ociv, the exact factor
decrease is not well measured. This is due to low
number statistics of \civ\ absorbers at $z\sim6$
together with differences in resolution and
sensitivity across different studies. 
Interestingly, this drop in \civ\ seems to be balanced with a rise in \cii\ \citep{cooper2019}.

The decline in the number of \civ\ absorbers with redshift
 has been attributed to a combination of 
a decrease in metallicity --the metal content of the CGM-IGM-- towards
higher redshift and a decrease in the intensity of the
background ionizing flux
that such metals are exposed to.
This change could be caused by a reduction, towards larger redshift,
in the number of sources with detectable \civ\ systems 
and/or a change in the covering fraction of the absorbing gas.
Both effects are probably in place.
In the first case, the number of sources 
with detectable \civ\ would decrease
because high ionization regions are less uniform
and less common towards the epoch of reionization.
In the second case, the assosiation to young galaxies implies 
recent start formation,
thus outlows are in their 
early stages and
will produce streams of baryons 
with a small projected area on the sky. 

Models support both interpretations:
\cite{finlator2015} cite a
combination of ongoing
enrichment and ionization 
in roughly equal measure. 
By exmaining the evolution 
of low ionziation systems 
the degeneracy between metallicity and UVB can 
be broken. Hints of an 
increase in the number of 
low equivalant width \mgii\
systems is seen around a redhisft of 5 
\citep{bosman2017, chen2017, codoreanu2017}. 
Evidence of the effect on the softening 
of the UVB on CGM scales has
been recently demonstrated 
by \oi\ absorption lines 
showing a significant upturn 
in their number density at
$z>5.7$ \citep{becker2019}. 

Although these arguments needs to be tested with larger
samples, our findings of faint galaxies
in close proximity to \civ\ systems at $z>5$
suggest that the absorbing material was ejected 
in the recent past. Figure \ref{f:ew-dist-wind}
shows that for \civ\ systems 7, 8 and 9,
the LAEs are within the reach of outflows
of $\langle v \rangle \sim 100$--150\,\kms\ starting at $z=10$. 
This is evidence that we are witnessing the first ejection 
of metals to the CGM-IGM, which would indicate that
one of the reasons of the drop in \Ociv\ is 
a decrease in the amount of metals outside of galaxies,
reflected in the covering fraction.

\section{Conclusion}\label{s:conclusion}

We conducted a search for LAEs 
in the line of sight to QSO J1030+0524,
using MUSE at VLT, to identify
the galaxies within $\simlt 250$ pkpc 
(transversal distance) 
from of 11 \civ\ absorptions systems
in the redshift range $z=4.76$--5.97.
The results can be summarized as follows:

\begin{itemize}
\item We report the detection of
at least one LAE in close 
proximity to the four \civ\ systems with 
$\log_{10}(N_{\text{\civ}}[$\cm$])>13.5$
and multiple LAEs for the systems at $z_{\text{\civ}}>5.5$.
The closest LAEs
are also within $\pm250$\,\kms\ from 
the corresponding \civ. The exception 
are LAE $\#7$ and $\#8$ at $-639$\,\kms from \civ\ 9.
In contrast, we find no LAE within $\rho< 200$--250 pkpc
of the seven weaker \civ\ systems  
($\log_{10}(N_{\text{\civ}}[$\cm$])<13.5$),
although two of them have one
LAE neighbor at 335 pkpc.
Overall, \civ\ systems with $\log_{10}(N_{\text{\civ}}[$\cm$])>13.5$
at $z=5$--6 are more likely to have galaxies with \Lya\ emission 
within $\rho< 200$ pkpc (4/4 cases) than the \civ\ systems 
with $\log_{10}(N_{\text{\civ}}[$\cm$])<13.5$
 (0/7 cases). 

\item The absence of LAE candidates
for seven \civ\ systems with 
$\log_{10}(N_{\text{\civ}}[$\cm$])<13.5$
\AA\ is difficult to be explained by strong 
outflows from galaxies at $\rho> 200$ pkpc
(outside MUSE's FoV).
We conclude that the true sources
of weak \civ\ absorbing clouds
remain below the detection limit
of this work ($ 2\times 10^{-18}$ erg s$^{-1}$ \cm)
because their \Lya\ luminosity is simply too low,
or they have stopped emitting \Lya\ photons, 
or the emission is absorbed by intervening
neutral hydrogen (LLSs, DLAs, etc).

\item We present LAE \#6, which is a 
 0.28\,L$^{\star}_{\text{\Lya}}$
galaxy at $\rho=10$ pkpc from the strongest \civ,
and represents the first clear example
of enrichment by satellites
and neighbor dwarf galaxies at very early times.
LAE \#5 is the brightest LAE in the sample
(1.15\,L$^{\star}_{\text{\Lya}}$)
which was detected with
standard broad-band images
by \citet{diaz2014}.
LAE \#5 and LAE \#6 
are detected at opposite sides of the QSO's line-of-sight
separated by $\sim140$ pkpc in projected distance,
and $\sim 50$\,\kms\ in velocity.
Thus, the two galaxies could be 
gravitationally bound.
This is 
observational
evidence that supports the prediction
from hydrodynamical simulations
that low-mass satellites drive
the CGM-IGM enrichment
at very early times
\citep[e.g.][]{garcia2017a}.

\item The comparison of the W$_0$ vs. $\rho$ 
of individual galaxies
shows that the \civ\ in absorption 
at $z>4.94$ in the sight-line of this study 
is similar to the CGM of low-$z$ dwarf galaxies
in low-density environments.
This is another indication
that the systems at $z>4.94$ are tracing the
early contribution of low luminosity and low mass
galaxies to the metal content of the CGM-IGM.

\item 
We find that the environment of \civ\ systems 
within 200 pkpc 
is populated by the faint end of 
the \Lya\ luminosity function
with some galaxies reaching luminosities 
fainter than 0.2 $L^{\star}(\Lya)$
(0.5 dex fainter than those currently 
used to determine the luminosity function)
and M$_{\text{UV}}<-20.5$ mag.
This is an indication that \civ\ at these redshifts are tracing
gas in the proximities of young and small galaxies,
similar to the currently best candidates
to drive the epoch of reionization at $z>6$. 

\item The detection of several low luminosity LAEs
among the highest redshift \civ,
and the lack of bright counterparts for 11 \civ\ systems
could indicate a connection between environment
and the ionization state of the CGM-IGM.
For example, if \civ\ systems at $z\,\simgt\,5.5$
are sensitive to the local ionizing conditions provided by the LAEs,
it might be possible to use \civ\ systems to trace 
similar regions at higher redshift in the search
for highly ionizing sources.

\item Our findings of faint galaxies
in close proximity to \civ\ systems at $z>5$
suggest that the absorbing material was ejected 
in the recent past. Figure \ref{f:ew-dist-wind}
shows that, except for LAE \#1,
all the closest pairs are below the 200\,\kms\ dotted line.
Also, for \civ\ systems 7, 8 and 9,
the LAE counterpart is within the reach of outflows
of $\langle v \rangle \sim 100$--150\,\kms\ starting at $z=10$. 
\end{itemize}

The connection between faint star-forming galaxies
and high-ionization absorption systems reported in this work,
is potentially a consequence of the role of satellites in the early
evolution of the CGM and the IGM.
Assuming that faint LAEs 
are mainly young star-forming galaxies,
the detection of these galaxies near strong \civ\ systems
supports the scenario in which
the decrease in the \civ\ comoving mass density
\Ociv\ and the column density distribution function
of \civ\ towards $z>5.3$
is a combination of:
a) low covering fraction 
due to a decrease in the amount of metals outside of galaxies,
and b) a bias with environment 
possibly driven by
the effect of local ionizing sources
in the UVB.

The mechanical feedback of outflows
required to eject metals to the CGM-IGM,
might also have an impact on the reionization of 
hydrogen in the IGM
by pouring holes in the star-forming regions
which allow the escape of ionizing (Lyman continuum) photons.
As a result, the same internal processes
that would connect LAEs and \civ\ systems
would also provide the means for the escape of ionizing photons.
In this picture, the same faint low-mass galaxies providers of metals 
to the high-$z$ IGM are also important sources of ionizing photons.
 
Finally, this work clearly demonstrates the key role of large 
integral field units like MUSE in the search for very 
faint emission line galaxies.
The MUSE Hubble Ultra Deep Field 
Survey has demonstrated the detection of low luminosity 
galaxies whose only signature is a faint \Lya\
emission line, that is, no continuum is detected, including
 in deep HST images \citep[see e.g.][]{hashimoto2017}. 
 JWST will not capture these faint Lyman-alpha emitting 
 galaxies with its imaging surveys and ground-based 
 spectroscopy can only follow-up the space-based detections. 
 Our data demonstrates that the \Lya\ 
 emission of the faintest galaxies is only a few Angstrom wide, 
 which is less than one pixel of the JWST NIRCam GRISM (1 nm/pixel). 
However, MUSE wavelength coverage ends at 930nm 
corresponding to \Lya\ $z\sim6.6$. 
Proposed future instruments such as the 
Keck Cosmic Reionziation Mapper (KCRM) will continue to 
1050nm (\Lya\ at $z=7.6$). This will allow the detection of 
faint Lyman-alpha emitting galaxies 
between redshift 6.6 and 7.6
which is critical to understanding the reionization of the Universe by 
 (1) measuring the attenuation of \Lya\ emission equivalent 
 width with respect to the underlying UV continuum \citep[e.g.][]{furusawa2016} 
 and (2) a census of the galaxies most likely to have the largest contribution to 
 the UV radiation that reionized the Universe at $z>6$.

\section*{Acknowledgements}

CGD would like to thank Kristian Finlator for all the interesting discussions 
that helped improving this work.
CGD acknowledges the support of the 
Consejo de Investigaciones Cient\'ificas y T\'ecnicas (CONICET) and the Gemini Observatory.
ERW acknowledges that parts of this research were conducted 
by the Australian Research Council Centre of Excellence 
for All Sky Astrophysics in 3 Dimensions (ASTRO 3D), 
through project number CE170100013.
KIC acknowledges funding from the European Research Council through 
the award of the Consolidator Grant ID 681627-BUILDUP.
SS acknowledges support from the ERC Starting Grant
NEFERTITI H2020/804240.

\bibliographystyle{mnras}
\bibliography{gdiazcivmuse} 

\appendix

\section{Sources identified in previous studies in the field of view.} \label{app:im_others}

\begin{figure*}
	\includegraphics[width=\textwidth]{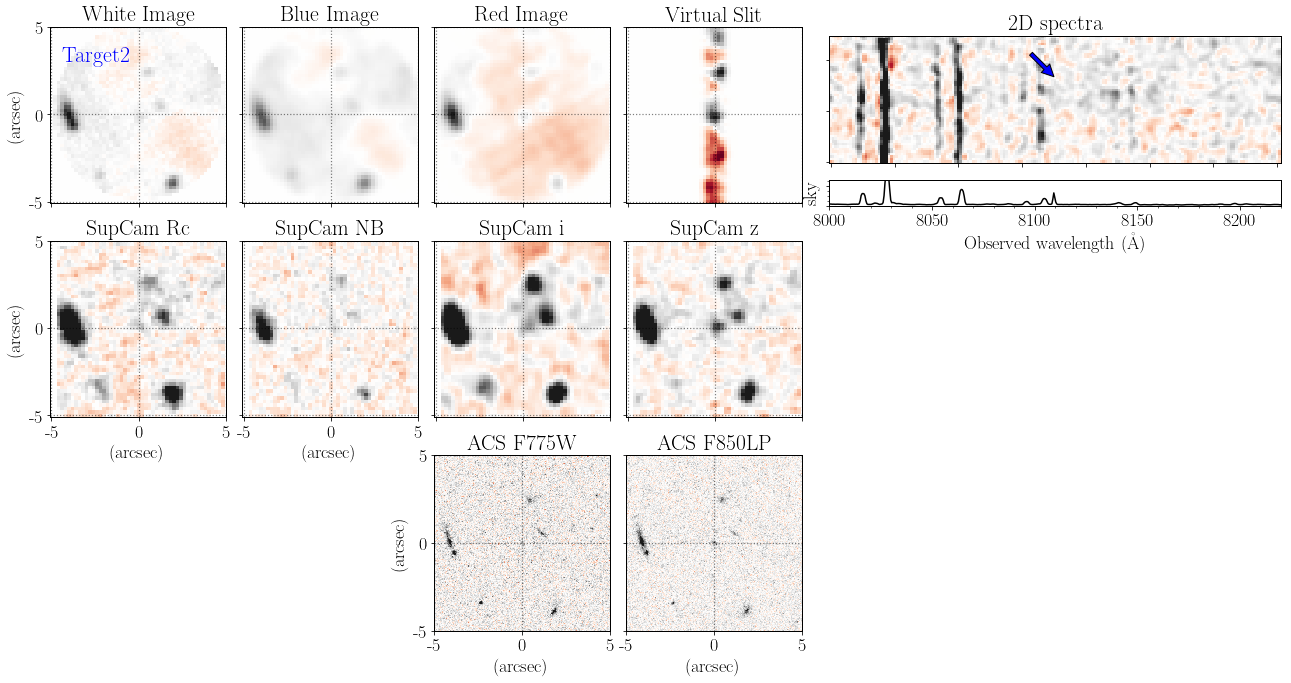}
	\caption{\small Target 2 from \citet{diaz2011}. 
		{\bf Top row:} MUSE thumbnails of $10\times10$\arcsec 
		centered on the object. Blue and Red are broad-band images
		of  1000\,\kms\ bandwidth at wavelengths bluer and redder  
		than rest-frame \Lya.
		The virtual slit used for the extraction of the 2D spectra is
		shown using a narrow band image for \Lya\ at $z_{em}=5.676$
		The 2D spectrum shows no evidence of emission line.
		{\bf Middle row:} SuprimeCam images from \citet{diaz2014}.
		{\bf Bottom row:} ACS HST images
		in i'-band (F775W) and z'-band (F850LP) from which the object was originally
		identified \citep{stiavelli2005}.}
	\label{f:target2}
\end{figure*}

\begin{figure*}
	\includegraphics[width=\textwidth]{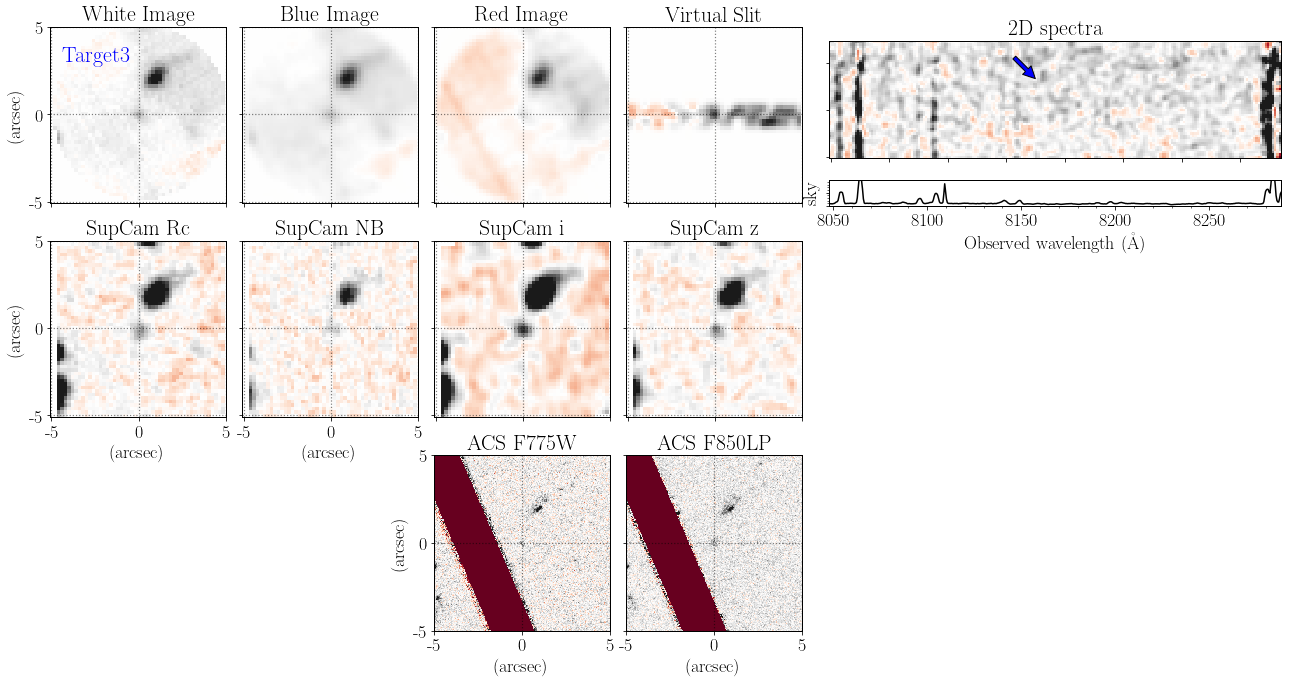}
	\caption{\small Target 3 from \citet{diaz2011}. 
		Description as per Figure \ref{f:target2}.}
	\label{f:target3}
\end{figure*}

\begin{figure*}
	\includegraphics[scale=0.5]{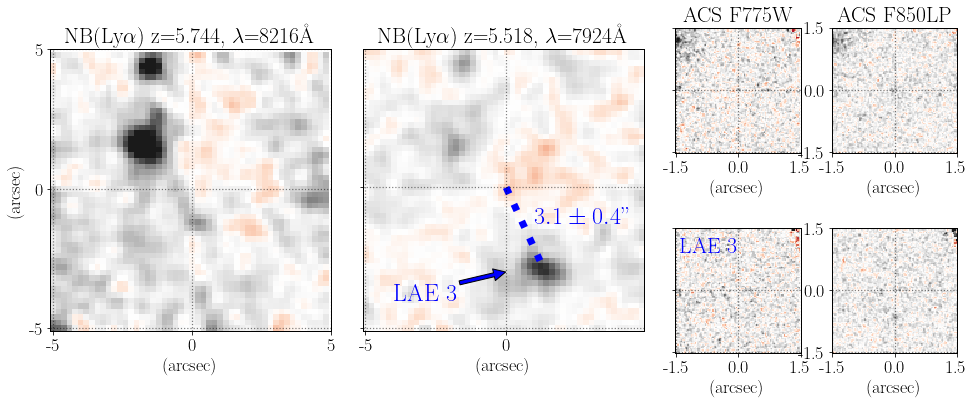}
	\caption{\small Object 1 from \citet{cai2017}. 
        {\bf Left and center:} MUSE thumbnails of
        $10\times10$\arcsec narrow band images for
        rest-frame \Lya\ at $z_{em}=5.744$ and
        $z_{em}=5.518$
		centered on the object's coordinates.
		{\bf Right:} ACS HST images
		in i'-band (F775W) and z'-band (F850LP) 
		at the position of Object 1 from
		\citet{cai2017} (top) and LAE $\#3$.}
	\label{f:cai1}
\end{figure*}

\begin{figure}
	\includegraphics[scale=0.3]{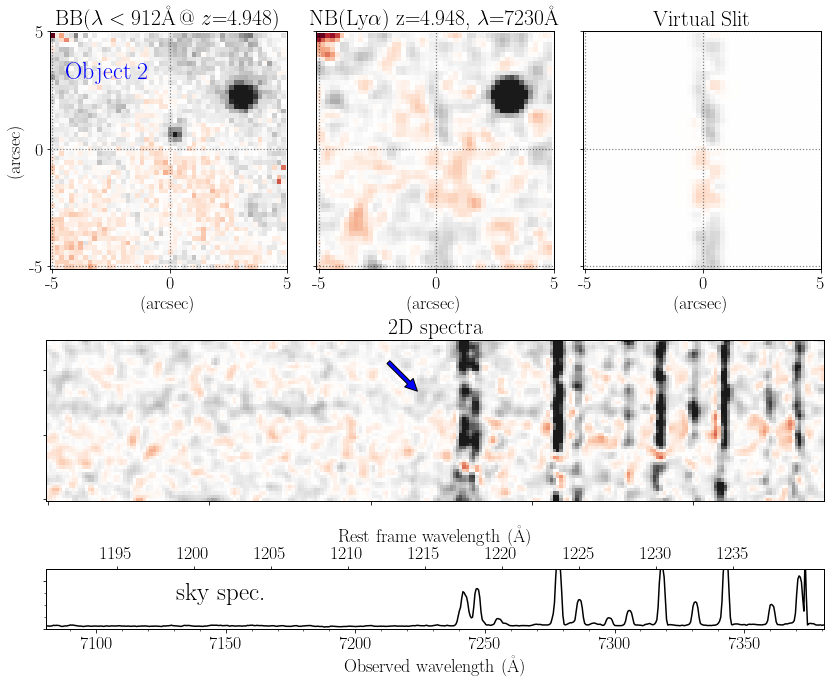}
	\caption{\small Object 2 from \citet{cai2017}. 
			{\bf Top row:} MUSE thumbnails of $10\times10$\arcsec 
		centered on the object. 
		{\it Left}: BB is a broad-band image
		including the all wavelengths 
		bluer than rest-fram $\lambda = 912$\,\AA\
		at $z=4.948$. This image reveals 
		a foreground
		object slightly avobe the center of 
		the field. 
		{\it Center}: Narrow band image for \Lya\ at $z_{em}=5.948$. No signal is detected.
		{\it Right}: Virtual slit used for the extraction of the 2D spectrum.
		{\bf Middle row:}
		The 2D spectrum shows no evidence of emission line. However, the continuum
		of the foreground object
		is detected across the wavelength range
		of the image.}
	\label{f:cai2}
\end{figure}

\bsp	
\label{lastpage}
\end{document}